\documentclass[10pt,aps,prd,twocolumn,tightenlines,superscriptaddress, letterpaper, amsmath, amssymb, preprintnumbers, floatfix, longbibliography, nofootinbib]{revtex4-2}
\usepackage{graphicx}
\usepackage{amsmath}
\usepackage[dvipsnames]{xcolor}
\usepackage{wrapfig}
\usepackage{float}
\usepackage{tikz}
\usepackage{array}
\usepackage{qcircuit}
\usepackage{comment}
\usepackage{enumitem}
\usepackage{bm}
\usepackage{tabularx}
\usepackage{multirow}
\usepackage{soul}
\usepackage{orcidlink}
\usepackage{physics}
\usepackage{subcaption}
\usepackage{orcidlink}

\usepackage{xcolor}
\definecolor{verdescuro}{RGB}{0, 150, 50}

\begin{document}

\title{Quantum Computing for Industrial Electromagnetics: Applicability and Case Studies in Solving Maxwell's Equations}
 
\author{Francesco Turro \orcidlink{0000-0002-1107-2873} }\email{francesco.turro@leonardo.com}  %
\affiliation{Quantum Computing Solutions,  Leonardo S.p.A., Via R. Pieragostini 80, Genova, Italy}    

\author{Marco Maronese \orcidlink{0000-0001-5548-5947} }
\affiliation{Quantum Computing Solutions,  Leonardo S.p.A., Via R. Pieragostini 80, Genova, Italy}

\author{Daniele Dragoni '\orcidlink{0009-0009-9637-4771}}
\affiliation{Quantum Computing Solutions,  Leonardo S.p.A., Via R. Pieragostini 80, Genova, Italy}
\affiliation{Leonardo HyperComputing Continuum, Leonardo S.p.A., Via R. Pieragostini 80, Genova, Italy}

\begin{abstract}
Computational electromagnetics plays a central role in many industrial applications but often requires substantial computational resources, particularly, when fine spatial discretizations are needed. While classical approaches remain the standard, quantum computing offers the potential to accelerate large-scale simulations by encoding them with a limited number of qubits.\\
Here, we investigate the performance and resource scaling of the Harrow–Hassidim–Lloyd (HHL) and Quantum Singular Value Transformation (QSVT) algorithms for solving linear systems generated by the finite-difference time-domain (FDTD) method, a widely adopted numerical scheme for discretizing Maxwell’s equations. We benchmark their performance across representative industrial use cases, including radar propagation, lens simulations, and beamforming processes. Our results demonstrate the validity of the approaches, achieving state infidelities smaller than $2\,\cdot 10^{-3}$ with success probabilities greater than $10^{-3}$, compatible with practical quantum state sampling. Overall, we observe that the QSVT method consistently delivers higher accuracy.
\\
We further observe that the condition number of the linear matrix, a key factor governing the performance of quantum solvers, saturates as the number of spatial lattice points increases. This means that the spatial grid can be scaled up to realistic industrial dimensions without inflating the HHL or QSVT circuit depth due to ill-conditioned matrices.
\end{abstract}

\maketitle

\section{Introduction}

Accurate numerical simulations of Maxwell’s equations play a central role in many industrial applications, including antenna designing~\cite{wang2024review,iyalagha2025beamforming}, radar performance analysis~\cite{li2005radar,Shittu2025radarcross}, lens and metalens engineering~\cite{rosenfield2016optometry,fu2021metalenses,jeon2023recent}. All these applications are based on solving the Maxwell's equations, which, in a material medium, can be written as:
\begin{equation}
\begin{split}
\nabla \cdot \mathbf{H}=0 \qquad & \nabla \times \mathbf{H}= \left( J+\frac{\partial \mathbf{D}}{\partial t}\right) \\
\nabla \cdot \mathbf{E} =\frac{\rho}{\epsilon_0} \qquad  &  \nabla \times \mathbf{E}=-\frac{\partial \mathbf{B}}{\partial t}\\
\mathbf{D}=\epsilon_r \epsilon_0 \mathbf{E} +\mathbf{P} \qquad & \mathbf{H}=\frac{\mathbf{B}}{\mu_r \mu_0}+\mathbf{M}
\end{split}
\end{equation}
where $\mathbf{D}$, $\mathbf{E}$, $\mathbf{B}$ and $\mathbf{H}$ indicate the electric displacement, the electric, magnetic, and magnetizing fields, respectively, $\rho$ the electric charge density, $J$ the current density, $\epsilon_0$ and $\mu_0$ the vacuum permittivity and permeability, $\mu_r$  and $\epsilon_r$ the relative permittivity  and permeability of a medium.\\
Numerous numerical methods have been developed for the simulation of Maxwell’s equations~\cite{Sumithra_Thiripurasundari_2017}. Among them, finite element methods (FEM)~\cite{jin2015finite} and finite volume methods (FVM)~\cite{Shankar01011990} offer excellent geometric flexibility and high accuracy for problems involving complex boundaries, heterogeneous media, and multiphysics couplings. However, these advantages come at the cost of increased mathematical and computational complexity, as they typically require the solution of large sparse linear systems, which can become computationally demanding for large-scale three-dimensional problems. The method of moments (MoM)~\cite{harrington1993field}, based on the transformation of integral equations into matrix systems through weighted residual techniques, is especially effective for open-boundary radiation and scattering problems. Since only surfaces or interfaces need to be discretized, MoM can significantly reduce the dimensionality of the problem. Nevertheless, the resulting matrices are generally dense, leading to unfavorable memory scaling and high computational cost for electrically large structures. More recently, discontinuous Galerkin time-domain (DGTD) methods~\cite{TOBON2015374} have attracted considerable attention because they combine high-order accuracy with local conservation properties and excellent scalability on parallel computing architectures. Despite these advantages, DGTD methods typically involve a larger number of degrees of freedom and more elaborate numerical flux treatments. Among finite difference methods, the finite-difference time-domain (FDTD) method~\cite{yee1966numerical} remains one of the most widely used approaches for broadband and transient electromagnetic simulations. Owing to its conceptual simplicity, explicit time integration scheme, and straightforward parallelization, FDTD has been extensively adopted in industrial applications ranging from antenna design and electromagnetic compatibility to photonics and radar modeling. However, FDTD requires discretization of the entire computational domain, which can result in substantial memory consumption and computational cost, especially for electrically large problems or structures involving fine or curved geometric details.\\
This important limitation of the FDTD method, the memory scaling drawback, motivates the exploration of alternative computational paradigms, such as quantum computing. Indeed, quantum computers can encode an exponential number of spatial degrees of freedom into a linear number of qubits. Specifically, we can encode $L=L_x L_y L_z$ lattice points ($L_i$ number of lattice points per axis) for $N_t$ time steps using a number of qubits $n_s\sim \log_2(L N_t) \sim  \log_2(L)+\log_2(N_t)$. In addition, quantum algorithms naturally exploit quantum parallelism, enabling the simultaneous manipulation of large state vectors and offering the prospect of polynomial or exponential speedups. \\
For these reasons, in this work, we reformulate the Finite-Difference Time-Domain (FDTD) method into corresponding linear systems of equations that can be solved efficiently with several proposed quantum algorithms~\cite{morales2025quantumlinearsolverssurvey}. The most prominent is the Harrow-Hassidim-Lloyd (HHL) algorithm~\cite{hhl} and its several variants~\cite{zaman2023step,PhysRevLett.110.230501,dervovic2018quantumlinearsystemsalgorithms,morgan2024enhancedhybridhhlalgorithm,saito2021iterativeimprovementmethodhhl}, which achieves logarithmic scaling relative to the system's matrix dimension, provided certain constraints on sparsity and condition number are met. As an alternative, adiabatic quantum evolution~\cite{Somma2019Quantumalgorithms,Dong2022quantumlinear} encodes the solution into the ground state of a slowly varying Hamiltonian, allowing the system to evolve naturally toward the desired result. Quantum Singular Value Transformation (QSVT)~\cite{gilyen2019quantum,shaviner2025quantumsingularvaluetransformation} further generalizes and unifies these linear-algebraic approaches; it enables efficient matrix inversion, projection, and spectral transformations with robust error control. More recently, the Schrödingerisation approach has emerged~\cite{jin2024QuantumSimulationPDE,ma2024schrodingerizationbasedquantumcircuits}. This technique maps linear differential equations onto a Schr\"odinger-type equation, facilitating efficient simulation on quantum hardware.\\
In this work, we evaluate the HHL and QSVT algorithms for solving FDTD systems across industrially relevant applications, the radar signal propagation and convex lens design, together with their computational resource requirements, since the practical relevance of a quantum solution depends not only on its accuracy but also on how the required resources scale with the problem size and with the specific settings of the considered use cases. Solution accuracy is quantified through the quantum-state fidelity with respect to the exact solution of the linear system, whereas the measurement overhead is characterized through the algorithmic success probability, which determines the number of shots needed to collect sufficient statistics. Particular attention is given to the condition number $\kappa$ of the system matrix, defined as the ratio between its largest and smallest singular values. 
Since $\kappa$ provides the connection between the properties of the underlying linear systems and the quantum resources required by the two algorithms, it is essential to understand its scaling. Specifically, in HHL, $\kappa$ affects the spectral resolution required in the phase-estimation procedure and, consequently, the size of the clock register and the circuit depth. In QSVT-based solvers, it determines the polynomial degree required to approximate the inverse function and therefore directly influences the circuit depth.\\
Our analysis reveals that the condition number approaches a plateau as the spatial grid is enlarged, indicating that it depends primarily on the number of temporal evolution steps $N_t$ and on the maximum dielectric contrast present in the computational domain. This behavior is consistent with the numerical performance obtained for both quantum algorithms. The saturation of $\kappa$ is particularly advantageous, as it implies that the spatial discretization can be increased to industrially relevant scales without a corresponding growth in quantum resources arising from matrix ill-conditioning. Consequently, large-scale electromagnetic simulations may be considered without incurring the substantial resource penalties typically associated with poorly conditioned linear systems.\\
This work is organized as follows: Sec.~\ref{sec:method} presents the metodology of the work. Sec.~\ref{sec:results} discusses the obtained results. 
Sec.~\ref{sec:discussion} recaps the advantage and disadvantage about the FDTD-HHL and FDTD-QSVT methods and 
Sec.~\ref{sec:conclusion} presents the conclusions.

\section{Method \label{sec:method}}

This section presents the method. In particular, Sec.~\ref{sec:FDTD} reviews the implemented FDTD method. Sec.~\ref{sec:fromFDTDtoHHL} formulates the FDTD method as a linear system of equations. Sec.~\ref{sec:quantumlinearsolver} recaps the main aspects of the HHL and QSVT algorithms.

\subsection{FDTD setup\label{sec:FDTD}}
\begin{figure}[b!]
\centering
\includegraphics[width=0.5\linewidth]{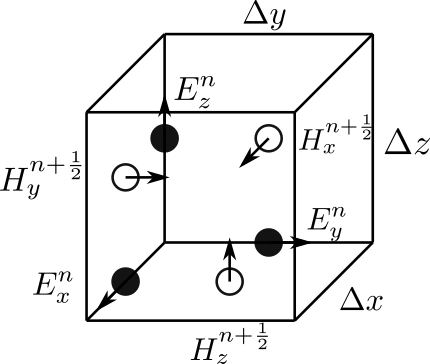}
\caption{Schematic picture of electric and magnetic fields in the Yee lattice formulation}
\label{fig:yeelattice}    
\end{figure}

Following the finite-difference time-domain (FDTD) method, we discretize space and time on a regular cubic grid with $(L_x,L_y,L_z)$ grid points, lattice spacings $\Delta \vec{s}=(\Delta x,\Delta y,\Delta z)$, and $N_t$ time steps of size $\Delta t$.
The electric and magnetic fields are discretized on a staggered spatial and temporal grid, known as the Yee lattice~\cite{yee1966numerical}. In this formulation, the components of the electric field are located on the edges of each computational cell, while the magnetic field components are positioned on the corresponding cell faces.
Let $\vec{n}=(i,j,k)$ denote a lattice point. For a discrete time index $t\in\{0,\,\ldots,\,N_t -1\}$, the field components are defined as
\begin{equation}
\begin{split}
    E_{\alpha}(t,n) &= \mathbf{E}_{\alpha}\left(t\,\Delta t,\left(\vec{n}+\frac{1}{2} \hat{\alpha}\right)
    \odot \Delta \vec{s} \right)\\
H_{\alpha}(t,n)&=\mathbf{H}_{\alpha}\left(t\,\Delta t-\frac{\Delta t}{2},\left(\vec{n}+\frac{\hat{x}+\hat{y}+\hat{z}-\hat{\alpha}}{2}\right) \odot \Delta \vec{s} \right)    
\end{split}
\end{equation}
where $\odot$ indicates the element-wise multiplication and $\alpha\in\{x,y,z\}$, $\hat{\alpha}$ is the corresponding Cartesian unit vector. Notice that there is a time shift between magnetic and electric fields\footnote{This time shift gives us the update equations for the centered finite differences with errors $O(\Delta t^2)$}.
A schematic representation of the field locations is shown in Fig.~\ref{fig:yeelattice}.
This staggered formulation preserves the discrete divergence constraints of Maxwell's equations, provided that they are satisfied by the initial conditions and by the source terms. 
Moreover, for each lattice point we define the electric permittivity and magnetic permeability constants for each field component, respectively, named $\epsilon(i,j,k)$ and $\mu(i,j,k)$. To address the case studied taken into account in this work we define a simplified FDTD setup where $\epsilon$ and $\mu$ are considered frequency independent and both electric and magnetic conductivity are set to $0$.\\
The update equations for electric and magnetic fields are obtained from the Maxwell's curl equations. All the FDTD update equations are reported in App.~\ref{app:FDTD_equations}.\\
We complete the specification of our FDTD setup by detailing the boundary conditions, the treatment of perfect electric conductors, the source implementation, and the stability criterion.\\
\textit{Boundary conditions.}
In simulations of realistic industrial use cases, the accurate modeling of wall boundary conditions is a crucial factor. Indeed, in radar applications or lens designing, it is essential to prevent electromagnetic fields reaching the computational boundaries from being reflected back into the interior of the simulation domain and measured from the detector. 
For this reason, we use the Perfectly Matched Layers (PML) boundary conditions~\cite{berenger1994perfectly} because they are specially designed to be absorbing layers that prevent artificial reflections of electromagnetic waves back into the simulation region. In particular, we adopt Berenger's classic split-field PML formulation~\cite{berenger1994perfectly}, wherein each field 
component is decomposed into two auxiliary components associated with the coordinate directions; for example, $E_z = E_{zx} + E_{zy}$ at grid points adjacent to the boundary (the PML region).%
This formulation ensures that waves entering the PML are attenuated and absorbed with minimal reflection, independent of their frequency and angle of incidence.\\

\textit{Metals.}
Metallic objects are modeled as perfect electric conductors (PECs), for which the tangential component of the electric field vanishes at the conductor surface, $\vec{S}\times \vec{E}=0$
$\vec{S}$ denotes the metal surface vector. In the FDTD implementation, this condition is enforced by setting all electric field components inside the PEC region to zero at every time step.\\

\textit{Electric sources.}
At the starting time $t=0$, we set both electric and magnetic fields to zero throughout the computational spatial domain. The generation of electric fields is done adding source terms $J$ into Amp\`ere's law:
\begin{equation}
  \nabla \times H = \epsilon \frac{\partial E}{\partial t}+J\,.
\end{equation}
Hence, rearranging, the electric-field update equation can be written as
\begin{equation}
 \frac{\partial E_{\alpha}}{\partial t}=  \frac{1}{\epsilon} \nabla \times H - \frac{1}{\epsilon} J_{\alpha} =  \frac{1}{\epsilon} \nabla \times H - S_{\alpha}(t,\,i,\,j,\,k)
\end{equation}
where $S_{\alpha}(t,\, i,\,j,\,k)$ denotes the source term applied in the FDTD grid at the point $(i,j,k)$ at the time $t$ for the $\alpha=x,y,z$ field components. In many applications, like radar design, the emitted beam is tailored in a specific direction. This is done using many sources and tuning their amplitude weights and phase shifts such that destructive interference suppresses radiation in undesired directions, while constructive interference powers the field in the desired direction. A description of time-shape of our sources and the beamforming process implementation can be found in App.~\ref{app:beamforming}.\\

\textit{Stability conditions.}
The FDTD method is numerically stable if we meet the Courant-Friedrichs-Lewy (CFL) Stability Conditions where the time step $\Delta t$ must be 
\begin{equation}
\Delta t \le \frac{1}{a_{CFL} c \sqrt{\frac{1}{\Delta x^2}+\frac{1}{\Delta y^2}+\frac{1}{\Delta z^2} } }\,.
\end{equation}
where $a_{CFL}\geq 1$ is a constant, in this work $a_{CFL}=\sqrt{3}$.\\
Moreover, as a rule‑of‑thumb~\cite{taflove2005computational}, the lattice spacing $\Delta\vec{s}$ should be $ \leq \lambda _{min}/10$, with $\lambda_{min}$ the minimum simulated wavelength to resolve correctly electromagnetic wave propagation.\\

\subsection{From FDTD equations to the FDTD Linear System of equations \label{sec:fromFDTDtoHHL}}

For solving the FDTD method with quantum computers implementing efficient quantum linear solvers, we have to reformulate the FDTD update equations in a linear system of equations of the form $A\,x=b$. This section presents this reformulation.\\
We define $\mathcal{E}(t)$ the vector containing all electric-field components in the Yee lattice,
\begin{equation}
\begin{split}
  \mathcal{E}(t) &= \left( 
E_x(t,0,0,0),\, E_y(t,0,0,0), E_z(t,0,0,0), \right. \\
& \qquad \left. ..., \, E_z(t, L_x-1,L_y-1,L_z-1) \right) ^T \,.
\end{split} 
\end{equation}
and $\mathcal{H}(t)$ is defined analogously for the magnetic-field components. We can hence define the vector $\mathbf{F}(t)$ as
\begin{equation}
\mathbf{F}(t)=
\begin{pmatrix}
\mathcal{E}(t),\, \, 
\mathcal{H}(t)
\end{pmatrix}^T,
\end{equation}
which encodes all electric and magnetic field components at time $t$.\\
Similarly, we define the source vector $\mathbf{s}_t$, which encodes the source terms at time $t$,
\begin{equation}
\begin{split}
  s_t &= \left( 
S_x(t,0,0,0),\, S_y(t,0,0,0), S_z(t,0,0,0), \right. \\
& \qquad \left. ..., \,S_z(t, L_x-1,L_y-1,L_z-1) \right) ^T \,.
\end{split} 
\end{equation}
We can formulate the FDTD equations for the vector $\mathbf{F}$ as it follows
\begin{equation}
\begin{split}
\mathbf{F}(t+\Delta t)&=
\begin{pmatrix}
1+C_H C_E & C_H\\
C_E & 1
\end{pmatrix}
\mathbf{F}(t)
+
\begin{pmatrix}
s_t\\
0
\end{pmatrix}\\
&= M\,\mathbf{F}(t) + \mathbf{s_t} \,,\label{eq:FDTD_evol_F}\end{split}
\end{equation}
where $C_E$ and $C_H$ are the matrices governing the curl operations for electric and magnetic fields, respectively, and $M$ denotes the full evolution matrix. A detailed description how we obtain Eq.~\eqref{eq:FDTD_evol_F} can be found in App.~\ref{app:linearequations}.\\
Following Refs.~\cite{li2024potentialquantumadvantagesimulation,turro}, the evolution of the electromagnetic fields from a general initial time step $t_s$ to $t_s+N_t$\footnote{In our notation, $t$, $t_s$ and $N_t$ indicate the discretized time step index, to compute the physical time we have to multiply to $\Delta t$}, encoded in the vector $x$,
\begin{equation}
x=
\left(
\mathbf{F}(t_s),
\mathbf{F}(t_s+1),
\ldots,
\mathbf{F}(t_s+N_t)
\right)^T,
\end{equation}
can be obtained by solving the linear system
\begin{equation}
\Tilde{A}\,x=b, \label{eq:Ax=b}
\end{equation}
where the matrix $\Tilde{A}$ and the right-hand-side vector $b$ are defined as
\begin{equation}
\Tilde{A}=
\begin{pmatrix}
1 & 0 & 0 & \cdots & 0\\
-M & 1 & 0 & \cdots & 0\\
0 & -M & 1 & \cdots & 0\\
\vdots & \vdots & \vdots & \ddots & \vdots\\
0 & 0 & 0 & \cdots & 1
\end{pmatrix} \qquad 
b=
\begin{pmatrix}
\mathbf{F}_0,\\
\mathbf{s}_{t_s},\\
\ldots,\\
\mathbf{s}_{t_s+N_t-1}
\end{pmatrix}\,.
\label{eq:TimeA}
\end{equation}

An extended mathematical description how we obtain Eq.~\eqref{eq:TimeA} can be found in App.~\ref{app:linearequations}.\\
The matrix $\tilde{A}$ is not Hermitian, which is a requirement of the HHL and QSVT quantum algorithms~\cite{hhl,gilyen2019quantum}.
However, by defining the matrix $A$ as
\begin{equation}
 A  = \begin{pmatrix}
   0 & \tilde{A} \\
   \tilde{A}^\dagger & 0 \\
 \end{pmatrix} \label{eq:A_HHL}.
\end{equation}
one obtains a Hermitian system that encodes the same FDTD dynamics. The corresponding linear system becomes
\begin{equation}
  A \begin{pmatrix}
    0 \\
    x(t)\\
    \end{pmatrix}
  =\begin{pmatrix}
    b \\
    0\\
  \end{pmatrix} =\vec{b}. 
\end{equation}
From this result, we can solve this linear system using quantum algorithms,
encoding the full matrix $A$ and vector $b$ using $n_{\rm sy}$ qubits, called system qubits. With this encoding, the computational cost of representing the spatial electric and magnetic field components scales logarithmically with the number of grid points and the cost of evolving the system over $N_t$ time time steps scales logarithmically with the number of qubits, requiring $n_{\rm sy} \sim \log_2(6 L_x L_y L_z (N_t + 1)) \sim \log_2(6 L_x L_y L_z)+\log_2(N_t + 1)$.
This logarithmic scaling offers significant advantages in both storage efficiency and the simulation of large lattices over extended time periods, as the required memory grows only logarithmically with system size. Consequently, this approach enables the simulation of later time steps with minimal quantum resource overhead, solving the main draback of the classical FDTD methods.

\subsection{Quantum linear solvers \label{sec:quantumlinearsolver}}

In order to solve the FDTD linear system, we consider two widely used quantum linear solvers: the Harrow-Hassidim-Lloyd (HHL) algorithm~\cite{hhl,zaman2023step,PhysRevLett.110.230501,dervovic2018quantumlinearsystemsalgorithms,tsemo2024enhancingharrowhassidimlloydhhlalgorithm} and the Quantum Singular Value Transformation (QSVT) framework~\cite{gilyen2019quantum}. The HHL algorithm encodes $b$ into a quantum state $\ket{b}$, performs quantum phase estimation to extract the eigenvalues of $A$, applies a controlled eigenvalue inversion, and uncomputes the eigenvalue register to prepare a state proportional to the solution $\ket{x}$. In contrast, QSVT-based linear solvers construct a quantum circuit that approximates with a $d$-degree polynomial function $f(A)\sim A^{-1}$, thereby, we have $\ket{x}= A^{-1} \ket{b} \sim f(A) \ket{b}$. For $d$-degree, the QSVT circuit is composed by implementing $d$ times a sequence of the block-encoding of $A$ with controlled phase-shift operators applied to extra ancilla qubit. These phases determine the accuracy of the solution of the linear system. A detailed description of implemented HHL and QSVT is reported in App.~\ref{app:linearsolvers}.\\
In the literature, the original time complexity of the Harrow-Hassidim-Lloyd (HHL) algorithm is given by $\mathcal{O}(\kappa^2 s^2 \log N / \epsilon)$~\cite{hhl}, where $N$ indicates the dimension of the matrix ($N \sim N_t L$ in our case) and $s$ is the matrix sparsity. Subsequent works have presented significant improvements over HHL~\cite{HHL_improvement1, PhysRevLett.120.050502}. In particular, the optimal scaling for dense matrices is given by $\mathcal{O}(\kappa^2 \|A\|_F \sqrt{N} \mathrm{polylog}(N) / \epsilon)$~\cite{PhysRevLett.120.050502}, where $\|\cdot\|_F$ denotes the Frobenius norm. In contrast, the quantum singular value transformation (QSVT) framework achieves a complexity scaling of $\mathcal{O}(\kappa \log(\kappa / \epsilon))$~\cite{gilyen2019quantum}.\\
The HHL algorithm requires a total number of qubits $n_{tot}=n_{\rm sy}+n_{\rm cl}+1$, where $n_{\rm cl}$ represents the number of clock qubits of the quantum phase estimation routine that sets the precision of the system solution. The number of clock qubit must higher than $n_{\rm cl} \geq \log_2(\kappa)$ where $\kappa$ indicates the condition number of $A$. In contrast, the QSVT approach requires $n_{\rm tot}=n_{\rm sy}+3$, independently of the target precision. In this case, one finds that the accuracy is controlled by the degree $d$ that grows approximately linearly with the condition number ($d\sim O(\kappa)$). \\
The successful execution of the HHL and QSVT algorithms relies on two crucial factors: the efficient encoding of the vector $\vec{b}$ on the system qubits,
and the accurate and resource-efficient realization of the required quantum circuits. In typical industrial applications, the vector $b$ solely encodes the source time functions, which are generally localized on a small subset of grid points. Consequently, $\vec{b}$ is highly sparse, with the vast majority of its components being zero. This localized structure enables an efficient state-preparation routine, drastically mitigating the gate complexity required to encode the right-hand-side vector onto the quantum register~\cite{ramacciotti2024simple, gleinig2021efficient}.\\
For the HHL algorithm, one finds that the total CNOT circuit depth is given by
\begin{equation}
  D_{HHL}\sim 2 n_{\rm cl} D_{HS}+2^{n_{\rm cl}}+n_{\rm cl}(n_{\rm cl}-1) \,,
\end{equation}
where $D_{HS}$ represents the depth of quantum circuit implementing a single controlled-$e^{iAt}$ operation, which typically provides the dominant contribution. The exponential factor comes from the Controlled-$Ry$ rotation sequence and $n_{\rm cl}(n_{\rm cl}-1)$ from the QFT circuit. However, the linear system matrix $A$ is highly sparse due to the locality of Maxwell's equations because each row contains at most seven non-zero elements, directly reflecting the local structure of Maxwell curl equations in the FDTD scheme. Explicitly, for a simulation spanning $N_t$ time steps across a spatial grid of $L$ lattice points, the matrix $A$ is expressed as:
\begin{equation}
A =\left. \frac{(\sigma_x+i\sigma_y)}{\sqrt{2}}\right|_{q_h} \otimes \left[ 1_{N_t}\otimes 1_{L} - S \otimes \sum_l^L m_l  \right] + h.c.\,\label{eq:A_paulis}
\end{equation}
where $S$ is a matrix with elements $S_{ij}=\delta_{i,j+1}$, $m_l$ denotes the localized Maxwell evolution operator acting on the $l$-th lattice point, $1_{k}$ represents the $k \times k$ identity matrix, and $\text{H.c.}$ denotes the Hermitian conjugate. The qubit $q_h$ is the qubit introduced to render the matrix self-adjoint (see Eq.~\eqref{eq:A_HHL}).\\
The Eq.~\eqref{eq:A_paulis} suggests that for the Hamiltonian simulation of the HHL method it can be decompose via Trotter decomposition. By exploiting this, the dynamics are approximated as a sequence of local propagators,
\begin{equation}
  e^{itA}\sim e^{i t \frac{\sigma_x+i\sigma_y}{\sqrt{2}}} \prod_{l,\tau} e^{-i t \frac{\sigma_x+i\sigma_y}{\sqrt{2}} \otimes \ket{\tau-1}\bra{\tau} \otimes m_l} ...
\end{equation}
where $\ket{\tau-1}\bra{\tau}$ operator selects the $\tau$ time step.\\
In contrast, the CNOT depth of the implemented QSVT linear solvers is given by
\begin{equation}
  D_{\rm QSVT} \sim \,d\,D_{A}\,,
\end{equation}
where $D_A$ corresponds to the depth of implementing the block encoding of $A$ whose compilation in gates can be nontrivial and may dominate the overall computational cost. For example, FABLE method~\cite{camps2022fable} incur a gate complexity of $\mathcal{O}(N^2)$, where $N$ is the dimension of $A$. One can exploit the sparsity of A by exploiting sparse position oracles where the block-encoding can be constructed with a gate complexity scaling as $\mathcal{O}({\rm polylog}(N))$~\cite{gilyen2019quantum}. Alternatively, we can also implement Linear Combination of Unitaries (LCU)~\cite{Childs20LCU} to decompose the block encoding matrix in quantum circuit, but the LCU success probability decays with the sum of coefficients that in our case is proportional to $L\,N_t$. \\
Moreover, another bottleneck of the QSVT algorithm is the optimization of the single controlled-phase gate that requires dedicated polynomial-synthesis procedures, that introduces a significant theoretical and computational complexity especially when $\kappa$ is high.\\
\begin{figure*}[t!]
\centering
\includegraphics[width=0.85\linewidth]{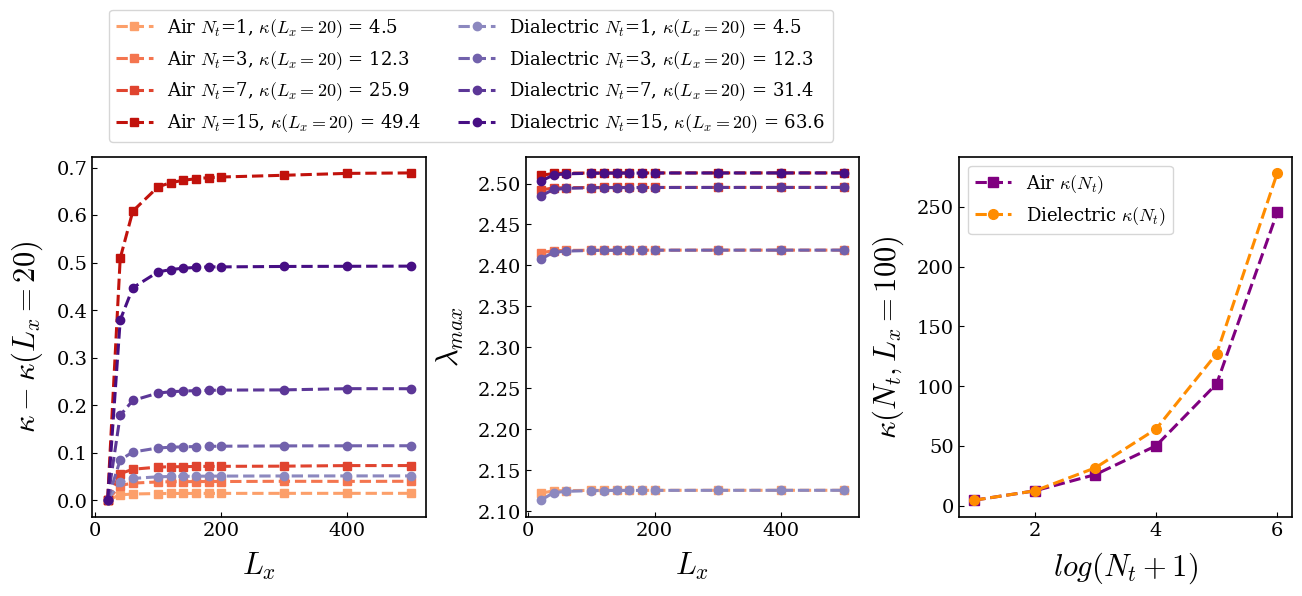}
\caption{Condition number $\kappa$ (left panel) and maximum eigenvalue $\lambda_{\max}$ (center panel) of the one-dimensional FDTD evolution matrix as functions of lattice size $L_x$ for different numbers of time steps $N_t=1,3,7,15$. Square markers denote propagation in air, while circles correspond to a configuration with a dielectric region of relative permittivity $\epsilon_r=4$ occupying the second half of the domain ($x \ge L_x/2$). Different colors corresponds to different evolution times. The right panel shows the condition number computed at $L_x=100$ as a function of $\log_2(N_t+1)$, specifically, the purple squares corresponds for the air configuration, dark orange circles for the dielectric system.}
\label{fig:conditionnumber1D}
\end{figure*}

\begin{figure*}[t!]
\centering
\includegraphics[width=0.85\linewidth]{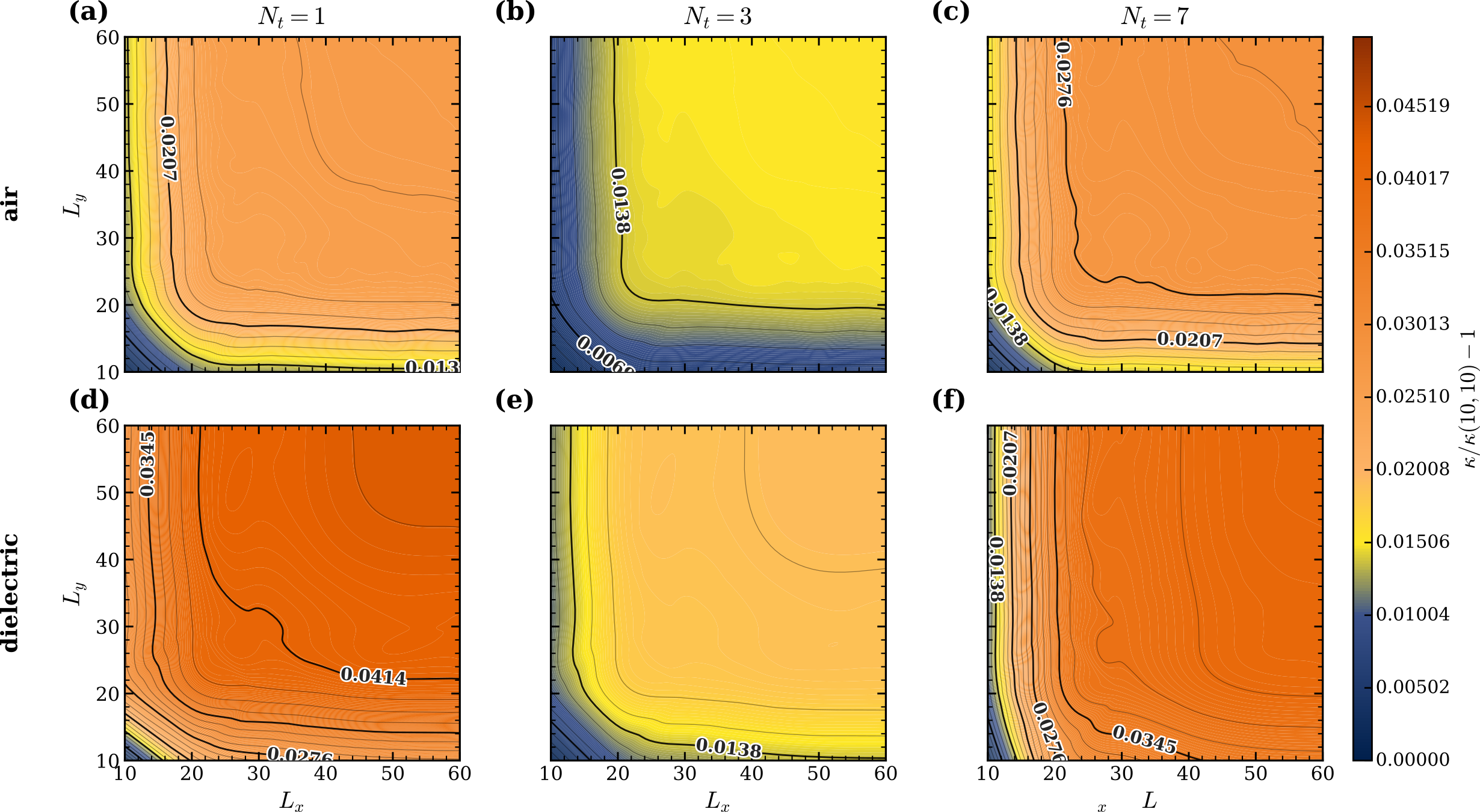}
\caption{Condition number $\kappa$ of the linear FDTD matrix as a function of the lattice sizes $L_x$ and $L_y$, for different evolving time steps $N_t = 1, 3, 7$ ( panel (a) and (d) for $N_t=1$, panel (b) and (e) for $N_t=3$,panel (c) and (f) for $N_t=7$). The (a),(b), (c) panels report the results correspond for the propagation in air, whereas the (d),(e), (f) panels show the results for a configuration in which the quadrant defined by $x \ge L_x/2$ and $y \ge L_y/2$ has relative permittivity $\epsilon_r = 4$, while the remaining region is filled with air.}
\label{fig:conditionnumber2D}
\end{figure*}

\section{Results \label{sec:results}}

The primary goal of this work is to assess the performance of quantum linear solvers in solving the FDTD linear systems, which can be quantified using two metrics.
The first metric is the success probability, $P_s$, which determines the expected number of shots required to extract the solution from the quantum hardware. For practical industrial applications, this probability should remain sufficiently large. In the HHL method, $P_s$ corresponds to the probability of successfully projecting the ancilla qubit into the $\ket{1}$ state and the clock register into the $\ket{0\dots0}$ state. For the QSVT-based solver, it represents the probability of measuring all ancilla qubits in the $\ket{0}$ state. More details on the number of shots can be found in App.~\ref{app:resource_estimation}.\\
The second metric assesses the accuracy of the quantum solver by comparing the exact solution of the linear system, $x$, with the state obtained by the quantum hardware after the successful post-selection, $\ket{\psi}$. This is quantified via the error fidelity, defined as
\begin{equation}
  \epsilon_F=1-\frac{\left|\bra{x}\ket{\psi}\right|^2}{\left|\bra{x}\ket{x}\right|\left|\bra{\psi}\ket{\psi}\right|} \,,
\end{equation}
where the $\ket{x}$ state encodes the vector of the exact solution of the linear system. \\
In terms of resources, we have already accounted for the number of qubits required by both solvers. Although circuit depth can be evaluated theoretically, it falls outside the scope of our current analysis. Therefore, we emulate the corresponding quantum circuits classically using the Davinci-1 HPC system~\cite{davinci1} without relying on explicit gate-level decompositions. For larger systems (characterized by a system register size $n_{\rm sy} \ge 14$), full exact matrix operations\footnote{For HHL, the Hamiltonian simulation, (controlled-$e^{itA}$) and for QSVT the application of block encoding} become computationally prohibitive. Consequently, we deploy the Lanczos method~\cite{lanczos1950iteration} to efficiently emulate the HHL and the QSVT circuits.
Conversely, for the numerical optimization of the QSVT angles, we utilize the open-source \texttt{qsppack MATLAB} package~\cite{qsppack_github,Dong2021qsppack,Wang2022energylandscapeof,Dong2024Robustiterative,Dong2024infinitequantum}.\\
In Sec.~\ref{sec:conditionnumber}, we analyze the condition number of the system matrix. We then discuss the performance of the HHL-FDTD and QSVT-FDTD algorithms across various applications. As an initial benchmark, we investigate a one-dimensional system with dielectric material (Sec.~\ref{sec:1D_case}) as a toy model that provides a detailed comparative study of the two methods. Next, we scale the evaluation to two industrial applications: radar signal propagation (Sec.~\ref{sec:radar_metal}) across two distinct aircraft geometries (one dielectric and one metallic) and an optical lens design problem (Sec.~\ref{sec:lens_results}).
\subsection{Spectra and condition number \label{sec:conditionnumber}}
We find that both the condition number $\kappa$ and the maximum eigenvalue $\lambda_{\max}$, two fundamental parameters governing the HHL and QSVT algorithms, remain essentially invariant as the number of spatial lattice points increases, depending solely on the number of time steps and the dielectric materials. This unique property significantly simplifies the configuration of the HHL solver for large-scale systems, as the required number of clock qubits and the total evolution time can be determined entirely independently of the spatial grid resolution and for the QSVT algorithm, we can fix the number of total single qubit angles. \\
Figure~\ref{fig:conditionnumber1D} shows the condition number $\kappa$ and the maximum eigenvalue $\lambda_{\max}$ of the linear FDTD evolution matrix for a one-dimensional grid as a function of the number of time steps. The left panel shows the deviation of the condition number computed at a given lattice size $L_x$ from the reference value obtained at $L_x=20$ as a function of $L_x$ for different time steps. The central panel displays 
$\lambda_{\max}$ for different numbers of time steps as a function of lattice size $L_x$. The right panel illustrates the dependence of the condition number, evaluated at $L_x=100$, as a function of the number of qubits that encodes the time steps. Interestingly, as reported in App.~\ref{app:conditionnumber_multilayers}, we find that the condition number for multi-dielectric layered systems is almost independent of the number of lattice points and depends mainly on the time step and the maximum relative dielectric constant.\\
Similarly, Fig.~\ref{fig:conditionnumber2D} extends the condition number analysis to two spatial dimensions. We compute the condition number for two systems: the first consisting entirely of air, and the second comprising air with a dielectric region ($\epsilon_r=4$) occupying one quadrant. Each panel of Fig.~\ref{fig:conditionnumber2D} shows the deviation of the condition number from a reference value (corresponding to $L_x=L_y=10$) for a specific evolution time step $N_t$. Top panels correspond to the condition number for the air system, the bottom ones for the dielectric systems.
We observe that the condition number exhibits clear plateaus as the lattice size increases, signaling convergence toward size-independent values in the large-system limit. We further vary the number of points in the PML region (i.e., the boundary layer), and observe the same qualitative behavior. We also find that the maximum eigenvalue of the linear system matrix is nearly independent of the number of lattice points and varies only slightly with the number of time steps. This analysis is reported in App.~\ref{app:lambdaMax} . \\
In conclusion, the observed saturation of the condition number $\kappa$ with increasing spatial lattice size constitutes a particularly favorable feature of the proposed formulation. This behavior indicates that the spatial discretization can be extended to industrially relevant scales without incurring a corresponding increase in the quantum resources required by either the HHL or QSVT algorithms due to matrix ill-conditioning. Furthermore, the asymptotic value of $\kappa$ can be accurately inferred from simulations performed on significantly smaller and computationally tractable grids. This observation substantially simplifies the resource-estimation procedure, enabling reliable predictions of quantum computational costs for large-scale electromagnetic simulations without requiring the explicit construction and analysis of the full system matrix.
\subsection{Evolution for a one-dimensional dielectric system\label{sec:1D_case}}
\begin{figure*}[!t]
\clearpage
\centering
\begin{subfigure}[!t]{0.99\linewidth}
\includegraphics[width=0.95\linewidth]{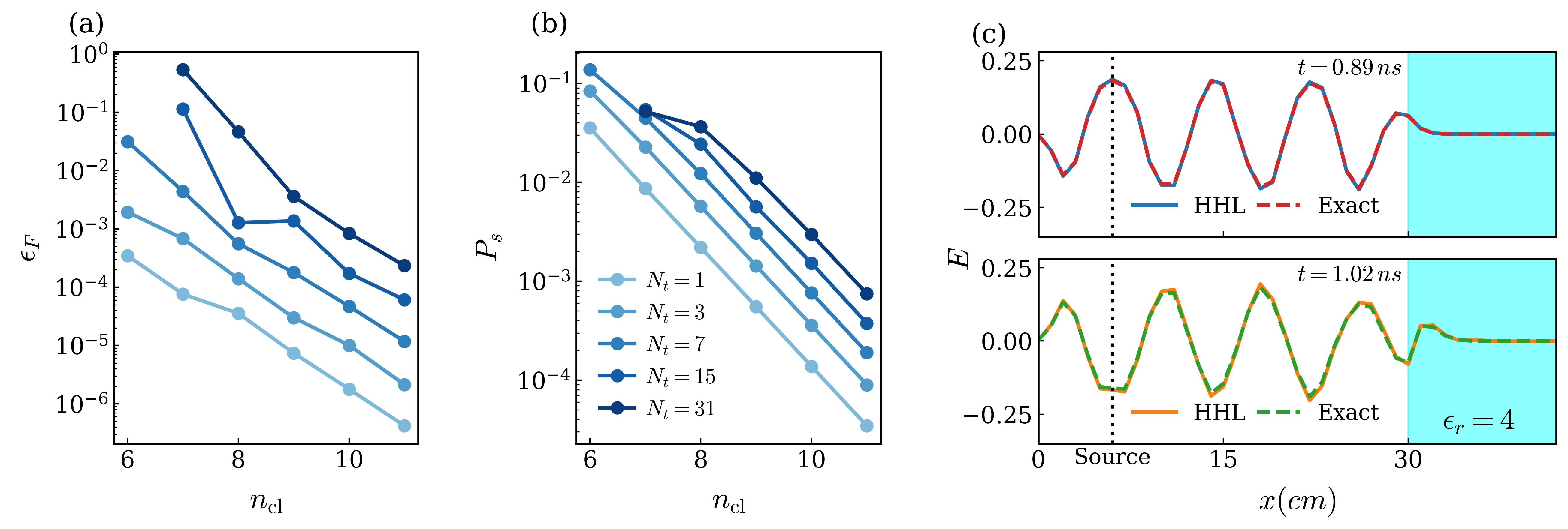}
\caption{Results of implementing the HHL-FDTD solver. 
The left and center panels show the fidelity error and the success probability as functions of the clock qubits for different number of time steps (indicated with different symbols). Right panel, obtained electric field for the HHL method for $n_{\rm cl}=8$ case. We could not implement HHL for the $n_{\rm cl}=6$ and $N_t=15$ case because $n_{\rm cl}<\log_2\kappa$.}
\label{fig:Evol_1D_Lx_42}
\end{subfigure}\\
\begin{subfigure}[t]{0.99\linewidth}
  \centering
  \includegraphics[width=0.95\linewidth]{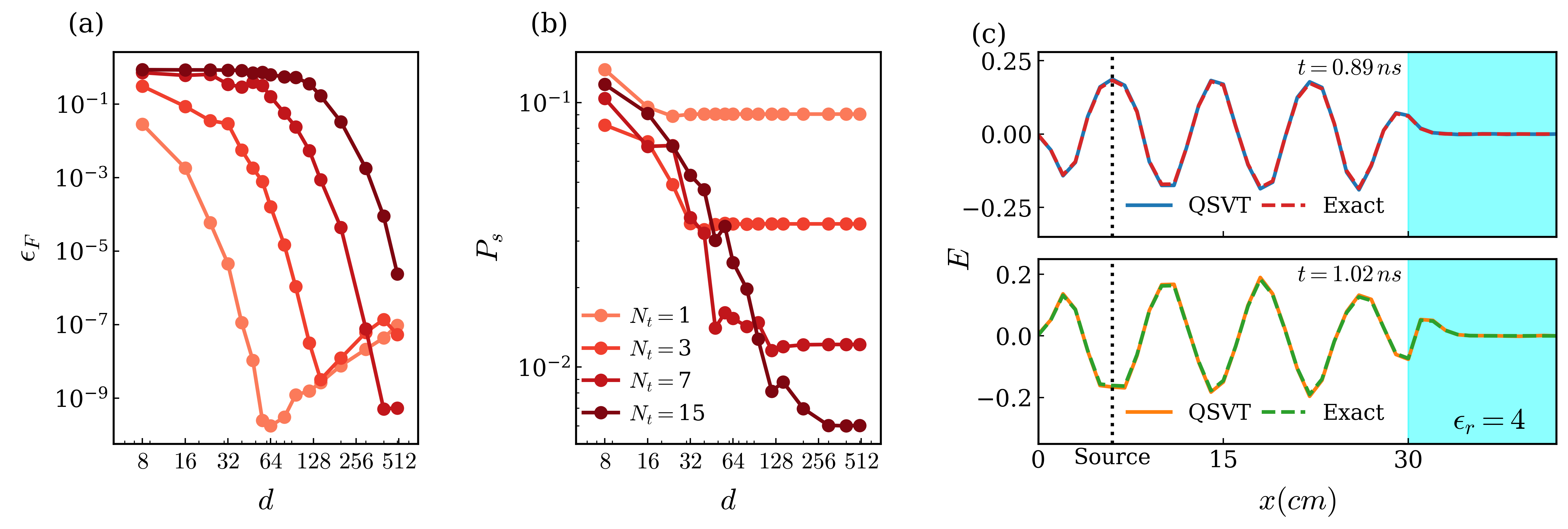} 
  \caption{Error fidelity (left panel) and success probability (right panel) of the FDTD-QSVT implementation for the benchmark system from Fig.~\ref{fig:Evol_1D_Lx_42}, plotted as a function of the number of QSVT angles $d$. Different colored symbols correspond to varying numbers of evolution time steps. Right panel, obtained electric field for the QSVT method for $d=400$. }
  \label{fig:QSVT_1D_results}
\end{subfigure}
\caption{Performance results of HHL-FDTD and QSVT-FDTD methods for a one-dimensional grid with $L_x=42$ lattice points, including a dielectric material placed from site 30 to the end of the chain. The upper and lower right panels in both subfigures show the obtained electric field using the two solvers for the $N_t=15$ case at $t = 80$ and $t = 92$, respectively, corresponding to when the wave enters the dielectric and after it has entered.
The dashed lines represent the analytical electric fields. The vertical dashed lines indicate the position of the source emitting a plane wave, while the cyan box denotes the dielectric material with $\epsilon_r = 4$.}
\end{figure*}
As an initial benchmark to assess the performance of the HHL-FDTD and QSVT-FDTD methods, we consider a one-dimensional grid of length $L_x = 42$, containing a dielectric material with relative permittivity $\epsilon_r = 4$ occupying sites 30 to the end of the chain. In our setup, a source term emits a plane wave with wavelength $\lambda = 8$. The simulations start from a time index $t_s = 80$ because the electric field starts scattering into the dielectric material.\\
We implement the HHL-FDTD method for different evolution time steps, $N_t = 1, 3, 7, 15, 31$, as a function of number of clock qubits, from $n_{\rm cl}=6$ to $11$ to assess its performance. The obtained fidelity error and success probability are reported in the left and center panels of Fig.~\ref{fig:Evol_1D_Lx_42}, respectively, where different colors correspond to different numbers of clock qubits. In the rightmost panels of Fig.~\ref{fig:Evol_1D_Lx_42}, the solid lines show the obtained electric fields at $t = 0.89 \,ns$ (top right panel) and $t = 1.02 \,ns$ (bottom right panel) for the case $n_{\rm cl} = 8$ and $N_t = 15$, while the dashed lines represent the exact evolution. Additionally, the vertical dashed line indicates the source position, and the cyan box highlights the dielectric region with $\epsilon_r = 4$.\\
Increasing the number of clock qubits improves the accuracy of the solution but simultaneously reduces the algorithm's success probability. We observe that larger values of $N_t$ require a correspondingly larger clock register to maintain a given error tolerance. This is also confirmed from the increase of the condition number with the time steps. Remarkably, fixing the clock qubits, we observe that the success probability increases with the number of time steps. For $N_t\leq 31$, our results suggest using $n_{\rm cl}=8$ because it provides an optimal trade-off between an accurate solution (error less than $1\%$) and a measurable success probability ($P_s>0.001$). App.~\ref{app:FDTD_HHL_tests} reports a further study where we vary the number of lattice points $L_x$, finding that performance variations remain minor, as predicted from the condition number analysis. \\
We perform a similar analysis for the FDTD-QSVT linear solver using the identical system, where we test the performance of the QSVT algorithm as a function of the polynomial degree $d$ (the number of QSVT angles). Figure~\ref{fig:QSVT_1D_results} displays the resulting error fidelity (left panel) and the success probability (right panel) versus $d$ for various time steps $N_t$, distinguished by colored symbols. As expected, increasing $d$ suppresses the solution error and leads to the convergence of the success probability toward an asymptotic value. This highlights a prominent advantage of the QSVT approach over the HHL algorithm: in HHL, mitigating the solution error generally incurs a strict penalty via a severe reduction in success probability. However, we note that beyond a certain threshold of $d$, the error fidelity for $N_t = 1$ and $N_t = 3$ increases slightly, reaching a floor of approximately $10^{-7}$. This behavior is attributed to errors in angle evaluation and accumulated numerical precision errors. Furthermore, the matrix condition number $\kappa$ significantly impacts the QSVT convergence. For lower values of $\kappa$, highly accurate solutions are achieved with relatively few QSVT angles. Conversely, as $\kappa$ scales up, a larger number of angles is required to achieve comparable precision, though the error continues to drop monotonically with increasing $d$.\\
A further analysis of the performance of our QSVT implementation when overestimating the true matrix condition number $\kappa$ during the QSVT phase-angle optimization process is presented in App.~\ref{app:QSVT_conditionnumber}. 
\subsection{Radar Signal Propagation}
\label{sec:radar_metal}
\begin{figure*}[t]
\centering
\includegraphics[width=0.95\linewidth]{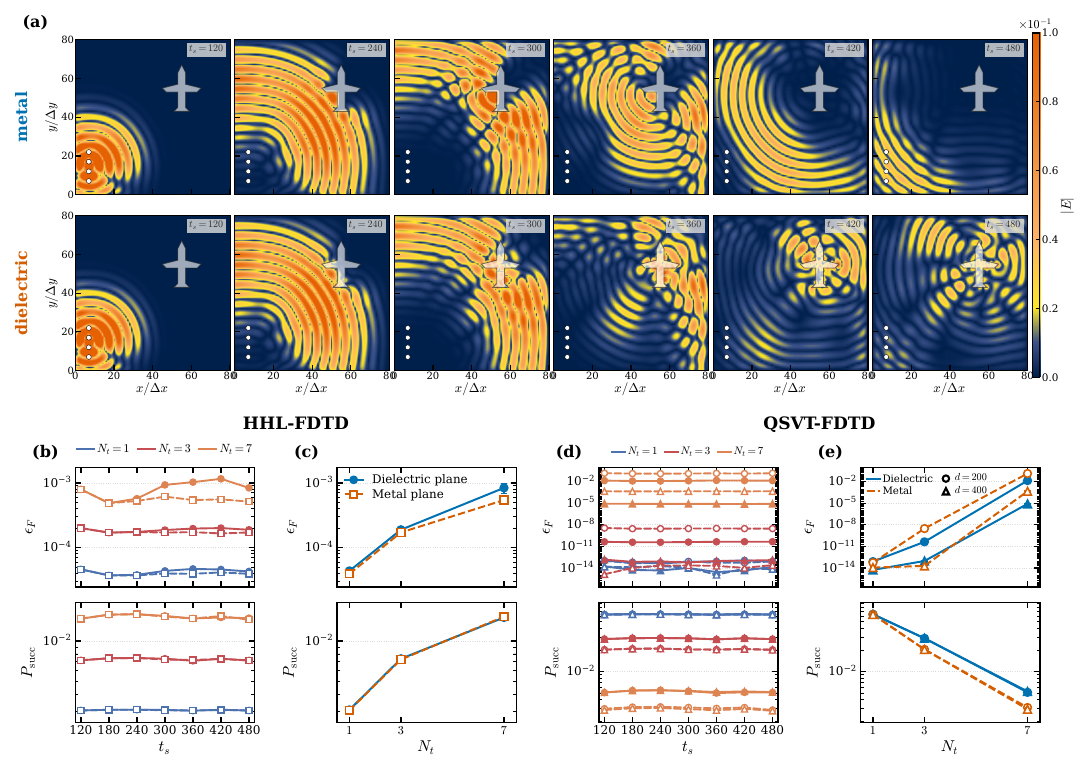}
\caption{Performance of the HHL-FDTD and QSVT-FDTD methods for the simulation of radar-signal propagation around two aircraft models: a metallic aircraft with perfect electric conductor (PEC) boundary conditions (diamonds) and a dielectric aircraft with relative permittivity $\epsilon_r=4$ (circles).
Panels (a) display the electric-field distributions for representative time steps. White circles indicate the source locations, while the gray polygons denote the aircraft geometry. In the upper field maps, the aircraft is made of metal; in the lower field maps, it is made of dielectric material.
Panel (b): Error fidelity (top) and success probability (bottom) of the HHL-FDTD method as functions of the initial time index $t_s$. Panel (c): Error fidelity (top) and success probability (bottom) of the HHL-FDTD method as functions of the number of time-evolution steps $N_t$. Panel (d): Error fidelity (top) and success probability (bottom) of the QSVT-FDTD method as functions of $t_s$ setting $d=200$ and $d=400$. Panel (e): Error fidelity (top) and success probability (bottom) of the QSVT-FDTD method as functions of $N_t$.
The two aircraft scenarios are reported with different line style, continuous line for dielectric aircraft, the dashed lines for metal. In panel (b) and (d) the colors indicate different evolution time steps, where different symbols represent the two different aircraft configurations. In panel (e), different symbols correspond to different solver configuration $d=200$ and $d=400$.}
\label{fig:radar_results}
\end{figure*}
A major, and undoubtedly relevant, industrial use case of electromagnetic simulation is radar signal propagation. In this scenario, sources emit a focused electromagnetic beam that propagates through the air and scatters off a target (typically an aircraft), enabling the precise measurement of the resulting scattered electric fields.\\
In our setup, we consider a simplified radar scenario on a two-dimensional grid of size $L_x=L_y=80$ containing a schematic aircraft profile, represented by the gray polygon in the field maps of panels (a) of Fig.~\ref{fig:radar_results}. Through a beamforming scheme, four sources (indicated by white circles) generate an electromagnetic beam propagating at an angle $\theta=\pi/6$ and with wavelength $\lambda=10$ (in lattice units). We consider two different target models: a dielectric aircraft with relative permittivity $\epsilon_r=4$ and a perfect electric conductor (PEC), representing a simplified metallic aircraft. The field maps in Fig.~\ref{fig:radar_results} illustrate representative stages of radar-signal propagation for both metallic and dielectric targets simulated for a single step ($N_t = 1$) using the QSVT solver with $d = 400$.\\
Panels (b) and (c) of Fig.~\ref{fig:radar_results} show the fidelity error and success probability of the HHL-FDTD approach as functions of the initial time index and the number of evolution steps ($N_t=1,\,3,\,7$), respectively, using $n_{\rm cl}=8$ clock qubits. The initial time index identifies different stages of the radar process, including signal emission, propagation through air, scattering from the target, and signal reception. In both panels, the upper plots report the fidelity error, while the lower plots show the corresponding success probability. Panels (d) and (e) present the analogous results for the QSVT-FDTD method with polynomial degrees $d=200$ and $d=400$.
From panels (b) and (d), both the fidelity error and the success probability exhibit only a weak dependence on the initial time index $t_s$ for both HHL-FDTD and QSVT-FDTD. This behavior suggests that the performance of the two algorithms is only weakly affected by the specific choice of the right-hand-side vector $b$ in the underlying linear system. Interestingly, the relative performance of the two solvers depends on the physical scenario: HHL yields slightly higher accuracy for the metallic-aircraft case, whereas QSVT achieves superior accuracy for the dielectric target.\\
As shown in panel (c), the HHL-FDTD fidelity error and success probability both increase with the evolution time step $N_t$. Nevertheless, all obtained errors remain below $10^{-3}$. For the QSVT-FDTD results shown in panel (e), the fidelity error decreases monotonically with increasing polynomial degree $d$, as expected. Although larger evolution times initially lead to higher baseline errors, these are systematically suppressed by increasing the polynomial degree. In particular, for $d=400$, the fidelity error falls below $10^{-3}$ for all considered configurations, while the success probability remains comparable to that obtained for $d=200$. We also observe a significant decrease in success probability with increasing $N_t$. Consequently, longer time evolutions require substantially more measurement shots to reconstruct the QSVT solution with comparable statistical accuracy.\\
In summary, both approaches provide highly accurate solutions, with errors below $10^{-3}$ and practically achievable success probabilities. For longer evolutions, however, the success probability decreases owing to the larger clock-register requirements of HHL and the unfavorable scaling of the QSVT success probability.
\subsection{Lens design\label{sec:lens_results}}

\begin{figure*}
\centering
\includegraphics[width=1.0\linewidth]{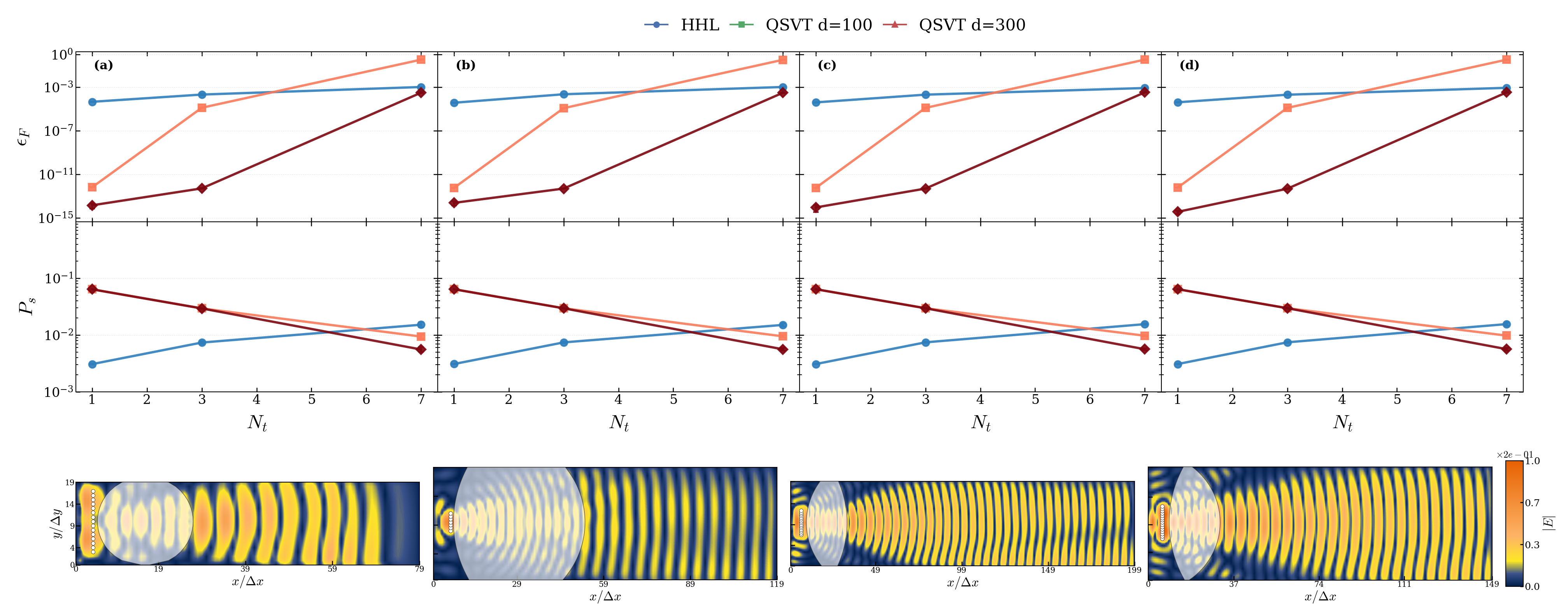}

\caption{Performance of the HHL-FDTD and QSVT-FDTD for simulating electromagnetic wave propagation through thin convex glass lenses. Each column corresponds to a different lens configuration that is shown in the bottom panels, where the colormap shows the electric fields at the studied times. Top panels display the error fidelity and middle ones the success probabilities. Blue curves indicate the HHL results obtained with $n_{\rm cl}=8$ and red ones the QSVT data setting $d=100$ and $d=300$. \label{fig:lens}}
\end{figure*}

We also consider the propagation of electromagnetic waves through a glassy lens. By time-averaging the squared electric field, the position of the maximum intensity identifies the focal point of the lens, useful in optimizing optical performance.\\
We consider several thin convex spherical glass lenses with refractive index $n=1.5$ to validate the proposed method, where each spherical lens is characterized by two curvature radii $R_1$ (right surface) and $R_2$ (left surface). The lens geometries and computational grids are shown in the bottom panels of Fig.~\ref{fig:lens}. Specifically, the lens and grid geometries for the different lens systems are shown in the bottom panels of Fig.~\ref{fig:lens}, where the lens of panel have $R_1 = R_2 =10$, panel (b) $R_1 = 40$, $R_2 = 30$, panel (c) $R_1 =40$, $R_2 = 60$, and panel (d) $R_1 = 50$, $R_2 =30$. Moreover, the white circles in the panels report the position of electric sources that emit in-phase spherical waves are placed before the lens, mimicking image-forming wave propagation.\\
Fig.~\ref{fig:lens} reports the fidelity error (top panels) and success probability (middle panels) obtained with the HHL-FDTD and QSVT-FDTD solvers for the different lens configurations as functions of the number of time steps $N_t$. Each column corresponds to the results obtained with the lens configuration shown in the bottom panels. The blue curves denote the HHL results obtained with $n_{\rm cl}=8$ clock qubits, whereas the red curves correspond to the QSVT results for polynomial degrees $d=100$ and $d=300$.
The same qualitative trends observed for the radar-scattering benchmarks are recovered in the lens simulations. For the HHL solver, the fidelity error increases with $N_t$, but remains below $10^{-3}$ in all configurations considered. The success probability increases with $N_t$. For the QSVT solver, increasing $N_t$ amplifies the baseline approximation error; however, this effect can be significantly mitigated by increasing the polynomial degree $d$. In contrast, the success probability remains largely insensitive to the choice of $d$, but it drops with $N_t$.\\
Consistent with condition number analysis, both the fidelity error and the success probability exhibit only minor variations across the different lens geometries. This behavior indicates that the performance of both HHL-FDTD and QSVT-FDTD is largely insensitive to changes in boundary geometry and spatial discretization. These results provide further evidence that the conditioning of the underlying linear system is governed predominantly by the temporal evolution length, rather than by local geometric features of the computational domain.

\section{Discussion\label{sec:discussion}}

Our results demonstrate that both approaches successfully solve the linear systems arising from the FDTD formulation. However, the QSVT-based method consistently achieves higher-fidelity solutions than HHL. Improving the accuracy of QSVT does not require additional qubits; instead, the circuit depth increases linearly with the polynomial degree $d$. In contrast, improving the accuracy of HHL requires additional clock qubits, resulting in deeper circuits and an exponential suppression of the success probability. Nevertheless, HHL benefits from simpler circuit decompositions compared to QSVT, even as the number of lattice points and time steps increases. For both methods, the preparation of the input vector $b$ is expected to be efficient because it encodes the temporal profile of a localized set of source terms\footnote{Assuming a zero-field configuration at $t=0$.}.

We find that the condition number of the linear system, which largely determines the resource requirements of quantum linear system solvers, exhibits only a weak dependence on the number of spatial grid points. This observation suggests that the asymptotic value of $\kappa$ can be accurately estimated from simulations performed on significantly smaller and classically tractable grids. Consequently, both approaches are expected to scale favorably toward industrially relevant problem sizes, without a corresponding increase in quantum resources due to matrix ill-conditioning. This favorable scaling is also reflected in the numerical performance of both algorithms.

In contrast, the condition number increases with the number of time steps, leading to a gradual degradation of algorithmic performance. For HHL, this degradation is primarily associated with the additional clock qubits, and hence a decay in success probabilit, required to maintain solution accuracy. For QSVT, our numerical results indicate that the degradation mainly arises from a reduction in success probability and from the increased difficulty of optimizing the QSVT phase angles. Furthermore, larger condition numbers generally lead to deeper quantum circuits. Nevertheless, for the systems considered here, the success probabilities remain sufficiently high for practical implementations, as the solution can be recovered with only a moderate sampling overhead.
To mitigate this performance degradation, quantum preconditioning techniques can be explored~\cite{nie2026paulistructuredpreconditioningquantumlinear,Clader2013Preconditioned,hosaka2024preconditioningvariationalquantumlinear}. By applying a sparse and easily invertible scaling matrix $M$, the linear system can be transformed into
$M^{-1}A x = M^{-1}b$,
while maintaining a reduced effective condition number. Such preconditioning strategies can improve the performance of both algorithms. Furthermore, the QSVT phase-angle optimization landscape is expected to become smoother, facilitating convergence.
\section{Conclusions \label{sec:conclusion}}

This work employs the finite-difference time-domain (FDTD) method to model the evolution of electric and magnetic fields at discrete time steps. By reformulating the FDTD method into a linear system of equations, we investigate the feasibility of simulating electromagnetic systems for industrial applications using two prominent quantum linear system algorithms (QLSAs): the Harrow-Hassidim-Lloyd (HHL) and Quantum Singular Value Transformation (QSVT) algorithms. Our results indicate a potential quantum advantage in representing spatial and temporal field variables. Specifically, the required qubit resources scale logarithmically with the number of discretization elements, effectively bypassing the primary memory bottlenecks of the classical FDTD method. Moreover, we also discuss that both the vector $b$ and the linear system matrix $A$ are extremely sparse, suggesting an efficient implementation in quantum circuits.\\
We show that the spectral properties of the resulting system matrices remain favorable as the spatial discretization is increased, with the condition number and largest eigenvalue approaching saturation. This behavior indicates that the quantum resources required by HHL and QSVT do not deteriorate with increasing spatial problem size due to ill-conditioning. Moreover, QSVT provides higher solution fidelity, while HHL offers simpler circuit implementations, highlighting complementary advantages of the two approaches.\\
Overall, our findings indicate that the proposed quantum methods constitute a promising framework for the simulation of large-scale electromagnetic systems. While significant challenges remain in the efficient realization of fault-tolerant quantum circuits, the favorable scaling observed in both the spectral properties and the memory requirements of the FDTD formulation provides encouraging evidence for the applicability of quantum computing to computational electromagnetics. Moreover, the favorable conditioning properties exhibited by the FDTD system matrices further strengthen this perspective, as they imply a resource scaling that remains manageable even for large problem instances. Taken together, these results identify the FDTD formulation as a particularly attractive candidate for quantum acceleration and establish a concrete pathway toward large-scale quantum computational electromagnetics. As fault-tolerant quantum technologies advance, these methods may unlock electromagnetic simulations beyond the reach of classical approaches, with potential applications spanning aerospace, defense, photonics, antenna engineering, and metamaterial design.

\bibliographystyle{unsrt}
\bibliography{./reference}

\newpage

\appendix

\section{FDTD update equations \label{app:FDTD_equations}}

All the update equations for electric and magnetic fields are obtained from the curl Maxwell's equations, specifically: 
\begin{widetext}
\begin{equation}
   H_x(t+1,i,j,k)=H_x(t,i,j,k)-\frac{\Delta t}{\mu(i,j,k) \mu_0} \left[ \frac{E_z(t,i,j+1,k)-E_z(t,i,j,k)}{\Delta y} -\frac{E_y(t,i,j,k+1)-E_y(t,i,j,k)}{\Delta z}\right ]\label{eq:H_x_FDTD}
\end{equation}

\begin{equation}
    H_y(t+1,i,j,k)=H_y(t,i,j,k)-\frac{\Delta t}{\mu(i,j,k)\mu_0} \left[ \frac{E_x(t,i,j,k+1)-E_x(t,i,j,k)}{\Delta z} - \frac{E_z(t,i+1,j,k)-E_z(t,i,j,k)}{\Delta x}\right ]\label{eq:H_y_FDTD}
\end{equation}

\begin{equation}
    H_z(t+1,i,j,k)=H_z(t,i,j,k)-\frac{\Delta t}{\mu(i,j,k)\mu_0} \left[ \frac{E_y(t,i+1,j,k)-E_y(t,i,j,k)}{\Delta x } -\frac{E_x(t,i,j+1,k)-E_x(t,i,j,k)}{\Delta y}\right ]\label{eq:H_z_FDTD}
\end{equation}

\begin{equation}
    E_x(t+1,i,j,k)=E_x(t,i,j,k)+\frac{\Delta t}{\epsilon(i,j,k)\epsilon_0} \left[ \frac{H_z(t+1,i,j,k)-H_z(t+1,i,j-1,k)}{\Delta y}-\frac{H_y(t+1,i,j,k)-H_y(t+1,i,j,k-1)}{\Delta z} \right ]\label{eq:E_x_FDTD}
\end{equation}

\begin{equation}
    E_y(t+1,i,j,k)=E_y(t,i,j,k)+\frac{\Delta t}{\epsilon(i,j,k)\epsilon_0} \left[ \frac{H_x(t+1,i,j,k)-H_x(t+1,i,j,k-1)}{\Delta z}-\frac{H_z(t+1,i,j,k)-H_z(t+1,i-1,j,k)}{\Delta x} \right ]\label{eq:E_y_FDTD}
\end{equation}

\begin{equation}
    E_z(t+1,i,j,k)=E_z(t,i,j,k)+\frac{\Delta t}{\epsilon(i,j,k)\epsilon_0} \left[ \frac{H_y(t+1,i,j,k)-H_y(t+1,i-1,j,k)}{\Delta x}-\frac{H_x(t+1,i,j,k)-H_x(t+1,i,j-1,k)}{\Delta y} \right ] \label{eq:E_z_FDTD}
\end{equation}
\end{widetext}
where $\epsilon(i,j,k)$ and $\mu(i,j,k)$ represent the values of dielectric and permeability functions in the position $(i,j,k)$. This setting is very general because the details of each simulation (objects, materials properties, ...) are only specified by setting the desired value of $\epsilon(i,k,k)$ and $\mu(i,j,k)$. In this work, for simplicity reasons, we consider lossless material (the electric and magnetic conductivity is set to $0$) and with constant $\epsilon_r$ and $\mu_r$ in the frequency spectrum.

\section{Our beamforming implementation\label{app:beamforming}}

\begin{figure}
\centering
\includegraphics[width=0.75\linewidth]{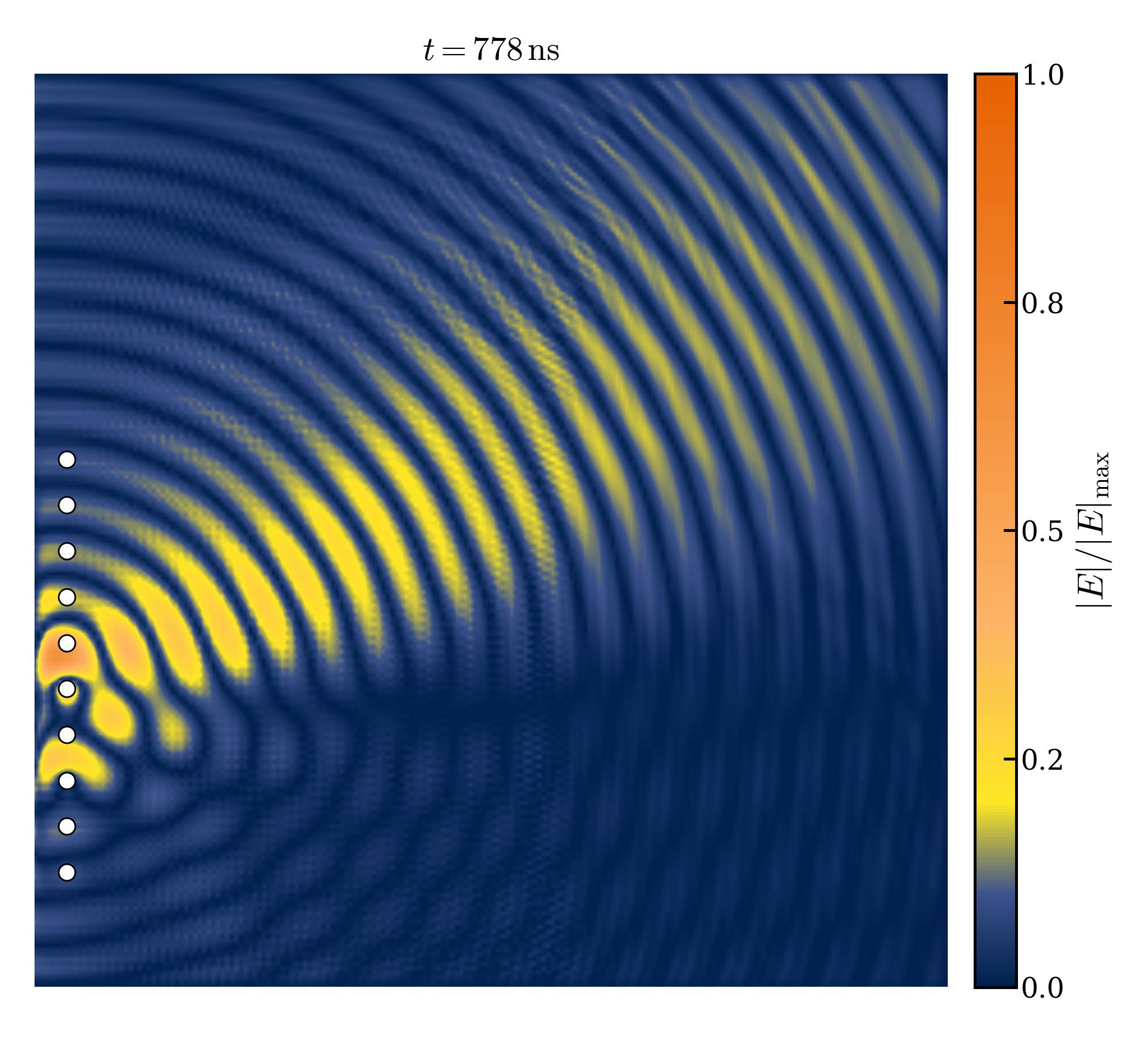}
\caption{Example of beamforming pattern using $N_s=10$ sources (indicated with white circles) with a direction angle $\theta=\pi/6$}
\label{fig:beamforming_example}
\end{figure}

As said in the main text, we use electric soft sources, which are independent of the local electric field and depend only on time 
$t$.\\
Specifically, we use the following source function \begin{equation}
    S_{\alpha}(t,\, i,\,j,\,k) = \frac{A(i,\,j,\,k)}{\sqrt{2 \pi}\Delta t} e^{ -\frac{(t - t_c)^2}{\sigma_t^2} } \sin( 2\pi \omega t + \phi_t )\label{eq:sourceshape}
\end{equation}
which describes a harmonic wave modulated by a Gaussian envelope and $A$, $t_c$, $\sigma_t$, $\omega$, and $\phi_t$ are the amplitude, temporal center, pulse width, angular frequency, and phase of the source, respectively.\\
In this work, the beamforming process is implemented for a two-dimensional grid using a one-dimensional source chain. Inside this chain, two adjacent sources are separated by $\lambda/2$, where $\lambda$ the emitted wavelength (that is computed from $\omega$).\\
We tailor the spatial amplitude distributions 
$A(n)$ in Eq.~\eqref{eq:sourceshape} with the Hamming window:
\begin{equation}
A_m= 0.54 - 0.46 \cos(\frac{2\pi m}{N_s-1}) 
\end{equation}
where $m$ indicates the $m$-th source out of $N_s$ sources. The beam steering angle $\theta$ is imposed through a progressive phase shift 
\begin{equation}
\phi_t^m= -\pi \,m\sin(\theta)
\end{equation}
Fig.~\ref{fig:beamforming_example} shows an example of the beamforming pattern.
\section{Linear solver \label{app:linearequations}}

In the main text, we define $\mathcal{E}$ and $\mathbf{B}$, the vector containing all the electric and magnetic fields. The FDTD curl equations for these two vectors can be written as
\begin{equation}
   \begin{split}
       \mathcal{H}(t+1)&=\mathcal{H}(t)+C_E \mathcal{E}(t)\\
       \mathcal{E}(t+1)&=\mathcal{E}(t)+C_H \mathcal{H}(t+1) +s_t \,,
   \end{split} \label{eq:FDTD_matrix_E_B}
\end{equation}
where the matrices $C_H$ and $C_E$ indicate the curl matrices for the electric and the magnetic fields. Attention must be paid to the order of solving these equations, first we evolve the magnetic field and then the electric one.\\
This last equation can be written as 
\begin{equation}
        \begin{pmatrix}
  E(t+1)\\
  B(t+1)\\
    \end{pmatrix}=\begin{pmatrix}
        1+C_H C_E & C_H\\
        C_E & 1\\
    \end{pmatrix} \begin{pmatrix}
  E(t)\\
  B(t)\\
    \end{pmatrix} + \begin{pmatrix}
    s_t\\
    0
    \end{pmatrix}
\end{equation}
We also define the vector $\mathbf{F}(t)$ as 
\begin{equation}
    \mathbf{F}(t)= \begin{pmatrix}
        \mathcal{E}(t)\,  \mathcal{H}(t)\\
    \end{pmatrix}^T
\end{equation}
the vector that encodes all electric and magnetic fields at time $t$. Eq.~\eqref{eq:FDTD_matrix_E_B} for the vector $\mathbf{F}$ can be written as 

\begin{equation}
    \mathbf{F}(t+1)=\begin{pmatrix}
        1+C_H C_E & C_H\\
        C_H & 1\\
    \end{pmatrix} \mathbf{F}(t) + \begin{pmatrix}
    \mathbf{s}_t\\
    0\\
    \end{pmatrix}
    =M\,\mathbf{F}(t) +
    \begin{pmatrix}
    \mathbf{s}_t\\
    0 \\
    \end{pmatrix} \,,
\end{equation}
where the $M$ matrix indicates the full matrix that evolves the electromagnetic fields according the FDTD equations.

Hence, the evolution of the electromagnetic fields, starting from an initial time $t_0=t_s \Delta t$ with the initial condition $F(t_s)=F_0$, can be written as:
\begin{equation}
\begin{split}
    \mathbf{F}(t_s) &= \mathbf{F}_0 \nonumber\\
    \mathbf{F}(t_s+1) - M \mathbf{F}(t_s)& =\mathbf{s}_{t_s} \nonumber\\
    ...&\nonumber\\
    \mathbf{F}(t_s+N_t) - M \mathbf{F}(t_s+(N_t-1))& =\mathbf{s}_{t_s+N_t-1}  \, .
\end{split}
\end{equation}
 where $s_t$ indicates the source vector at time $t$.
 
 We introduce the vector $x$ as:
\begin{equation}
    x=\left( \mathbf{F}(t_s), \mathbf{F}(t_s+1), ...,\mathbf{F}(t_s+N_t) \right) ^T \,.
\end{equation}
Following Ref.~\cite{li2024potentialquantumadvantagesimulation,turro}, we can write these equations in a compact linear system of form
\begin{equation}
\Tilde{A} x = b\,,
\end{equation}
where  the matrix $\Tilde{A}$ is defined as:
\begin{equation}
\Tilde{A}=    \begin{pmatrix}
        1 & 0 & 0 &...&0\\
        -M & 1 & 0 &...&0\\
        0 & -M & 1 &...&0\\
        ... &  ... & ... &...&0\\
        ... & 0 & 0 &...&1\\
    \end{pmatrix}\,, 
    \end{equation}
and the right-hand side vector $b$ is given by:
\begin{equation}
b= \begin{pmatrix}
    \mathbf{F}_0, & \mathbf{s}_{t_s}, & ..., & \mathbf{s}_{t_s+N_t-1}\\
\end{pmatrix}^T\,. \end{equation}

\section{Quantum Linear solver \label{app:linearsolvers}}

\subsubsection{The HHL algorithm}

\begin{figure}[t]
\centering
\includegraphics[width=1\columnwidth]{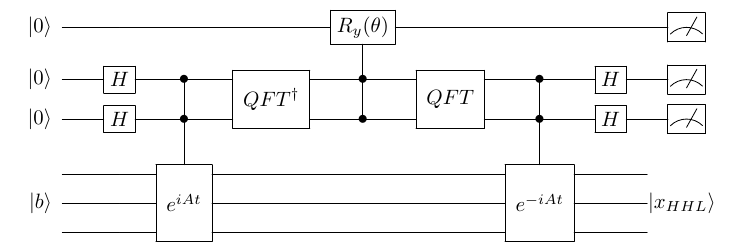}
\caption{Scheme of HHL circuit. The top qubit corresponds to the ancilla qubit, the second and third qubits represent the clock qubits, and the bottom register (fourth, fifth and sixth qubits) encodes the linear system in the system qubit. A detailed description of the single gates is reported in the main text.}
\label{fig:HHL_circuit}
\end{figure}

The HHL algorithm operates across three primary registers: an $n_b$-qubit system register encoding the linear system $A\vec{x}=\vec{b}$, an $n_c$-qubit clock register dictating numerical precision, and a single ancilla qubit. To avoid eigenvalue wrap-around, the clock register size must satisfy $n_c \geq \log_2\lceil \kappa \rceil$, where $\kappa$ is the condition number of $A$. As illustrated in Fig.~\ref{fig:HHL_circuit}, the protocol proceeds via five steps: we encode $b$ into the system register; then, we apply Quantum Phase Estimation (QPE)~\cite{Berry2006,Nielsen_Chuang_2010},  controlled-$R_y$ rotations targeting the ancilla qubit with the angles proportional to the inverse of the binary representation of the eigenvalue of $A$. In the last steps, we apply the inverse QPE and projective measurement, where the final solution of the linear system is found when the ancilla is measured in $\ket{1}$ state and the clock register in $\ket{0}^{\otimes n_c}$.\\
In this work, the time-evolution parameters in the QPE stage are set to be $t_i = 2^{n_i + 1} \pi / (4\lambda_{\max})$ for $n_i \in \{0, \dots, n_c-1\}$. This $1/4$ scaling factor maps positive and negative eigenvalues to distinct binary representations ($\lambda/4$ and $1 + \lambda/4$, respectively), successfully resolving degeneracies associated with $\pm\lambda_{\max}$ and ensuring robust eigenvalue separation.

\subsubsection{Quantum Singular Value Transformation}

\begin{figure*}
    \centering
    \includegraphics[width=0.9\textwidth]{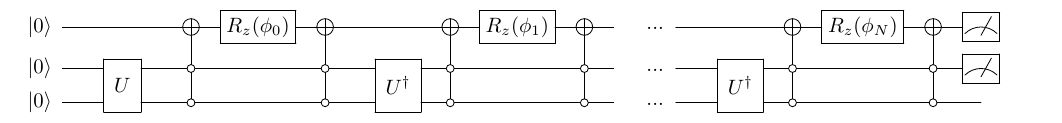}
    \caption{QSVT circuit for an even  number of angles.}
    \label{fig:QSVT_circuit}
\end{figure*}

Quantum Singular Value Transformation (QSVT) is a general quantum algorithm that enables the efficient implementation of polynomial transformation of the hermitian matrix $G$. QSVT can be also implemented for solving linear system of equations $A \ket{x} = \ket{b}$, where the primary objective is to compute the target state $\ket{x} = A^{-1} \ket{b}$. Within the QSVT framework, this is achieved by constructing a quantum circuit $C$ that implements a polynomial approximation of the inverse matrix, $C \approx f(A) \approx A^{-1}$. Specifically, we work with 
\begin{equation}
    A_1=A/\lambda_{\max}\,,
\end{equation} 
that is given by $A$ normalized with its eigenvalues,  and we are only interested in approximating $A_1^{-1}$ only up to the smallest singular value of the target matrix, that is given by the condition number $1/ \kappa$. 
Consequently, the solution is prepared directly as a quantum state: 
\begin{equation}
\ket{x}=A_1^{-1} \ket{b} \sim C \ket{b}\,,
\end{equation}
where the vector $b$ is encoded into a quantum state $\ket{b}$.\\
Fig.~\ref{fig:QSVT_circuit} illustrates the structural elements of a standard QSVT circuit. The core component is the unitary operator $U$, which represents the block-encoding of the matrix $A$. In our work, we use an extra ancilla qubit and the matrix A is embedded into the upper-left block of U such that:

\begin{equation}
U = \begin{pmatrix}
A_1 & i\sqrt{1-A_1^2}\\
i \sqrt{1-A_1^2} & A_1
\end{pmatrix}    \,.
\end{equation}
When the ancilla registers are measured and project onto the $\ket{0}$ ancilla state, the operation successfully applies $A$ to the system register (indicated with the lower qubits in the figure).\\
To transform the singular values of A into those of $A^{-1}$, the block-encoding U is interleaved with controlled phase-shift operators $e^{i\phi_i (2\ket{0}\bra{0} - I) }$ applied to extra ancilla qubit. The sequence of phase angles $\Phi =\left\{\phi_1,\phi_2,\dots,\phi_d\right\}$ dictates the specific polynomial transformation f(A). Crucially, these phase angles depend exclusively on the target mathematical function $f(x)\approx x^{-1}$ and the desired error tolerance, remaining completely independent of the structural details or specific entries of the matrix A itself. This universality is guaranteed by the foundational theorems of Quantum Signal Processing (QSP)\cite{Low2017QSP} and QSVT~\cite{gilyen2019quantum}.\\
In this work, we utilize the open-source \texttt{qsppack MATLAB} package~\cite{qsppack_github,Dong2021qsppack,Wang2022energylandscapeof,Dong2024Robustiterative,Dong2024infinitequantum} to numerically optimize and extract the phase angles $\phi_i$. We use an even number of angles, called $d$ in this work,  and $d$ dictates the precision of the reconstructed solution $\ket{x}$ like the number of clock in the HHL algorithm. Indeed, $d$ indicates the degree of polynomial that approximate $A^{-1}$, that, commonly,  are the Chebyshev polynomials. Generally, the algorithmic complexity requires a degree $d\sim\mathcal{O}(\kappa\log(1/\epsilon))$, where $\epsilon$ is the target accuracy. For practical implementations, a standard rule of thumb to achieve convergence across the entire non-zero spectrum is to choose a polynomial degree bounded by $d > 2 \kappa$.\\
Furthermore, we implement the LCU algorithm to select only the real of the obtained state, that is the solution.

\section{Shot estimation}
\label{app:resource_estimation}

In this work, we emulate with a state vector approach the HHL and QSVT FDTD solver, so, no finite-shot uncertainty is present. The reported values
of $P_s$ and $\epsilon_F$ are therefore deterministic quantities, up to algorithmic approximation errors, such as
Lanczos approximation errors, Hamiltonian-simulation errors, or QSVT
polynomial-approximation errors.

In contrast, in a shot-based simulation or on quantum hardware, the
statevector $\ket{\psi}$ is not directly accessible. The circuit is
executed $M$ times, and the success probability is estimated from the
number of successful post-selection events. If $M_{\rm succ}$ denotes the
number of successful shots, then
\begin{equation}
    \hat{P}_s =
    \frac{M_{\rm succ}}{M}.
    \label{eq:ps_estimator}
\end{equation}
Since each shot either satisfies or does not satisfy the post-selection
condition, $\hat{P}_s$ is a binomial estimator. Its statistical uncertainty
is
\begin{equation}
    \Delta P_s
    \simeq
    \sqrt{
    \frac{P_s(1-P_s)}{M}
    }.
    \label{eq:ps_uncertainty}
\end{equation}
Thus, estimating $P_s$ with additive precision $\delta_P$ requires
\begin{equation}
    M =
    O\left(
    \frac{1}{\delta_P^2}
    \right)
\end{equation}
shots in the worst case. However, only a fraction $P_s$ of all circuit
executions produces a valid post-selected solution sample. Therefore, the
expected number of total circuit executions required to obtain
$M_{\rm succ}$ successful samples is
\begin{equation}
    M
    \simeq
    \frac{M_{\rm succ}}{P_s}.
    \label{eq:shots_success_overhead}
\end{equation}
This relation shows that the success probability directly determines the
sampling overhead of the quantum solver.\\
In industrial applications, we desire to obtain the electric or magnetic fields. Therefore, one should perform quantum state tomography to obtain these values. For a system register of size
$n_{\rm sy}\sim \log_2(N_t L) $, the Hilbert-space dimension is
\begin{equation}
    d_{sy} = 2^{n_{\rm sy}}.
\end{equation}
A generic pure state contains $O(d_{sy} )$ independent real parameters. Hence,
the number of successful post-selected measurements required to reconstruct
a generic pure state with fidelity precision $\delta_F$ scales as
\begin{equation}
    M_{\rm succ}
    =
    O\left(
    \frac{2^{n_{\rm sy}}}{\delta_F^2}
    \right).
    \label{eq:tomography_successful_shots}
\end{equation}
Including the post-selection overhead, the total number of circuit
executions scales as
\begin{equation}
    M
    =
    O\left(
    \frac{2^{n_{\rm sy}}}{P_s \delta_F^2}
    \right) \sim O\left(
    \frac{T\,L}{P_s \delta_F^2}
    \right) .
    \label{eq:tomography_total_shots}
\end{equation}
This scaling is exponential in the number of system qubits and represents
the cost of reconstructing the full solution state from measurements. However, even though for larger grid the state tomography remains impracticable, the obtained scaling indicates the number of shots is linear in $T$ and $L$.

\section{Complexity dielectric structures \label{app:conditionnumber_multilayers}}

\begin{figure*}
\centering
 \includegraphics[width=0.95 \textwidth]{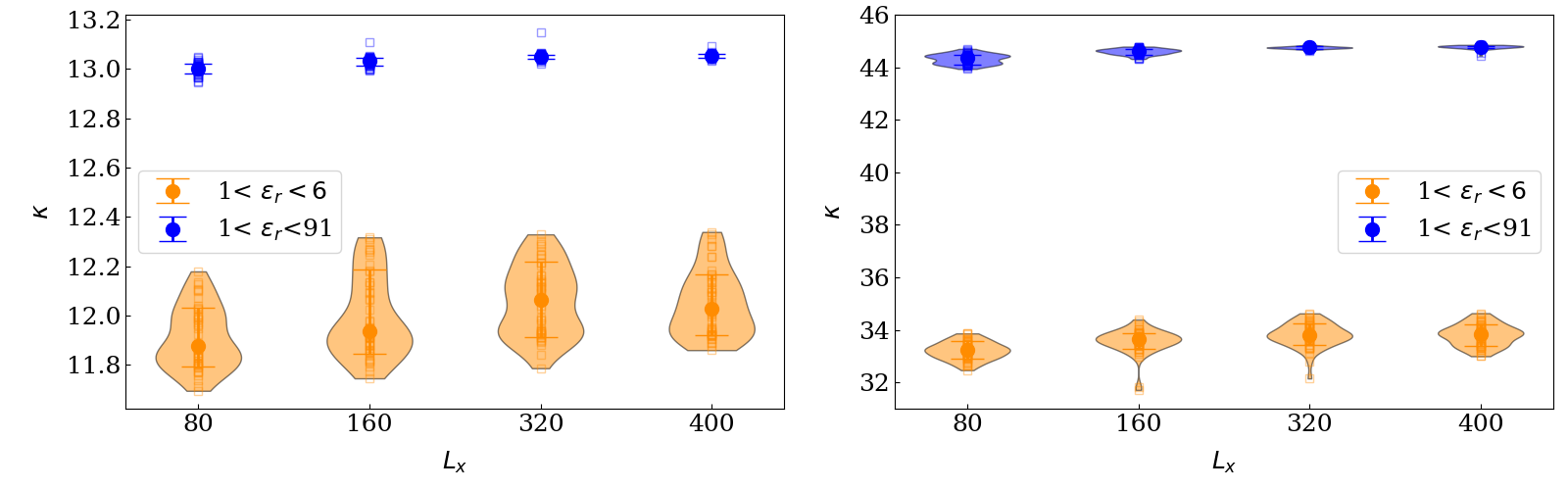}
\caption{Condition numbers for a complex dielectric configurations.
Specifically, different dielectric materials are placed every four lattice points, with randomly selected relative dielectric values in the ranges $1 < \epsilon_r < 6$ for orange squares and $1 < \epsilon_r < 91$ for blue squares. The condition number calculation is repeated over 50 random realizations for each system size. The violin plot in the left and right panels show the condition number results for the evolution for 3 time steps and for 7 time steps, respectively, as a function of lattice space $L_x$. 
Median and one-sigma variance are also reported with the error bars. }
\label{fig:dielectric_complex_1D}
\end{figure*}

In the main text, we show that, for a simplified system, the condition number of the evolution matrix depends only weakly on the number of lattice points for sufficiently large systems. In that one-dimensional grid analysis, however, the dielectric materials were placed continuously from the center of the system to the end of the grid. Here, we demonstrate that this behavior also holds for more complex configurations consisting of multiple dielectric structures. \\
Fig.~\ref{fig:dielectric_complex_1D} shows the obtained condition number results  for a 1-D grid for 50 different random configurations, where dielectric materials with a randomly selected relative dielectric constant in the range $1 < \epsilon_r < 6$ (orange) and $1 < \epsilon_r < 91$ (blue) are placed every four lattice points.\\
We observe that the condition number remains essentially independent of the number of lattice points. Our analysis indicates that $\kappa$ is primarily determined by the number of time steps and the maximum dielectric value present in the system. Furthermore, the condition number appears to be independent of the number or cumulative width of the dielectric layers. Notably, the condition number for these multilayer configurations is upper-bounded by the condition number obtained when a single block of the maximum dielectric value occupies exactly one-half of the 1D grid. 

\section{ Maximum eigenvalue for two-dimension systems\label{app:lambdaMax}}

\begin{figure*}[t!]
\centering
\includegraphics[width=0.9\textwidth]{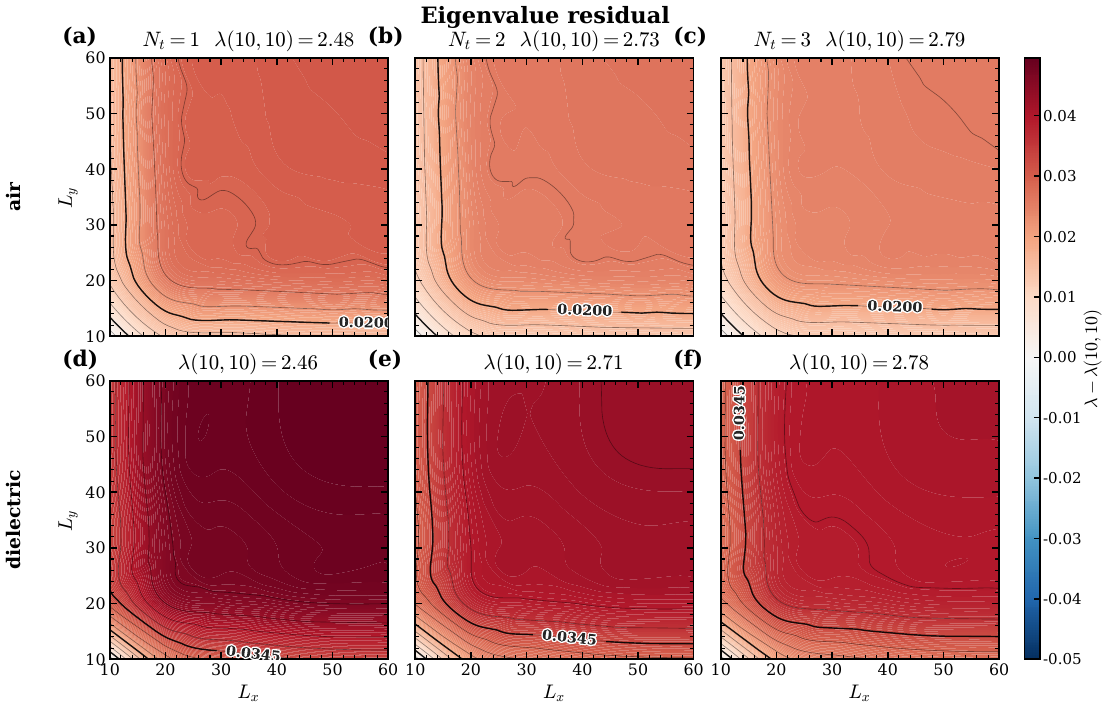}

\caption{Residual of the largest eigenvalue relative to the reference configuration at $(L_x=10,L_y=10)$ of the linear FDTD matrix, shown as a function of the lattice dimensions $L_x$ and $L_y$ for different numbers of time steps $N_t = 1, 3, 7$ (shown in different columns). The top panels correspond to propagation in air, whereas the bottom panels show results for a configuration in which the quadrant defined by $x \ge L_x/2$ and $y \ge L_y/2$ has relative permittivity $\epsilon_r = 4$, while the remaining region is filled with air.}
\label{fig:lambamax}
\end{figure*}

We analyze the maximum eigenvalue $\lambda_{\max}$ of the FDTD system matrix as a function of the lattice sizes $L_x$, $L_y$, and the number of time steps $N_t$. Fig.~\ref{fig:lambamax} shows the obtained results for the two systems considered in the main text as function of lattice grid, for different time steps. We find that $\lambda_{\max}$ is nearly independent of the spatial discretization and varies only weakly with $N_t$.

\section{Test of HHL-FDTD and QSVT FDTD methods on lattice size\label{app:FDTD_HHL_tests}}

\begin{figure*}
\centering
\includegraphics[width=1.0\linewidth]{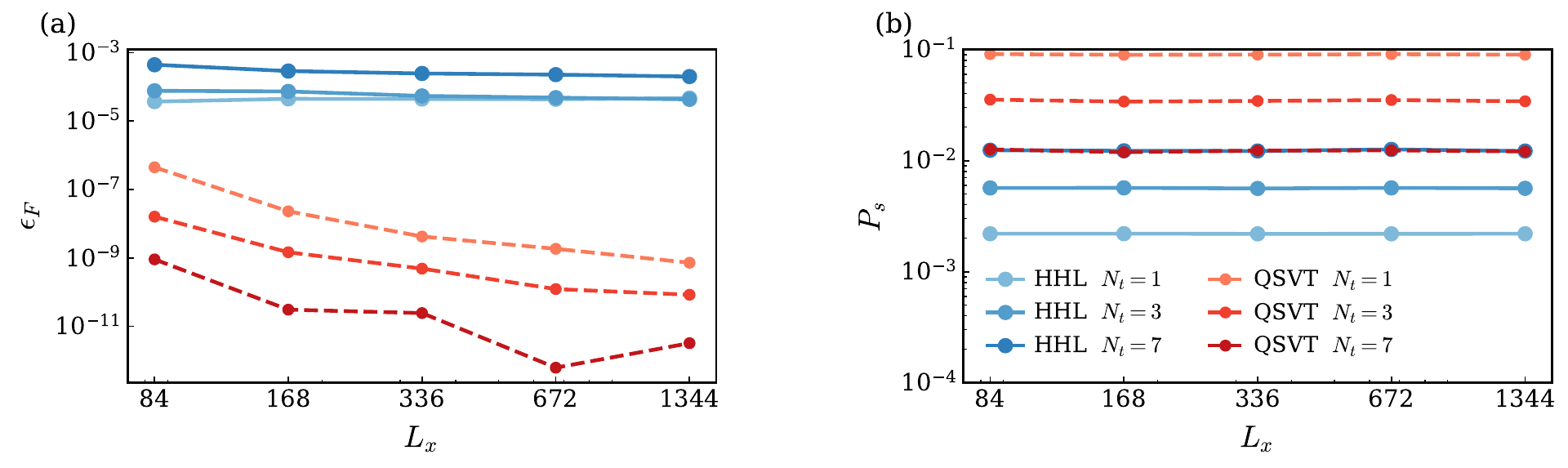}
\caption{Fidelity error (panel (a)) and success probability (panel (b)) as functions of the lattice size $L_x$ for the HHL- and QSVT-FDTD solvers in a one-dimensional grid. The dielectric medium has relative permittivity $\epsilon_r=4$ and occupies the sites from $x\in [L_x/2+8, L_x]$. Different colored markers correspond to $N_t=1,3,7$ time-evolution steps. HHL results are shown in blue with $n_{\rm cl}=8$, while QSVT results are shown in red with $d=500$.}
\label{fig:Evol_1D_Lx_size}
\end{figure*}

This section extends the test of the HHL and QSVT FDTD methods as the number of lattice points $L_x$ increases. We report the fidelity error and success probability, for one-dimensional grids with different $L_x$, where a dielectric material with relative constant $\epsilon_r=4$ occupies sites from $L_x/2 + 8$ to the end of the chain. Panels (a) and (b) of Fig.~\ref{fig:Evol_1D_Lx_size} show the fidelity error and success probability, respectively, as functions of the lattice size $L_x$ for different numbers of time-evolution steps, $N_t=1,\,3,\,7$. Results obtained with the HHL solver are shown as blue markers using $n_{\rm cl}=8$, while those obtained with the QSVT solver are shown as red markers with $d=500$.  The starting time index is scaled as $t_s = 80\,\frac{L_x}{42}$ to account for the grid size. We observe that, as $L_x$ increases, the fidelity error remains constant or decreases, while the success probability is largely independent of $L_x$.

\section{QSVT Performance Sensitivity to Condition Number Estimation \label{app:QSVT_conditionnumber} }

\begin{figure*}[t]
\centering
\includegraphics[width=0.7\linewidth]{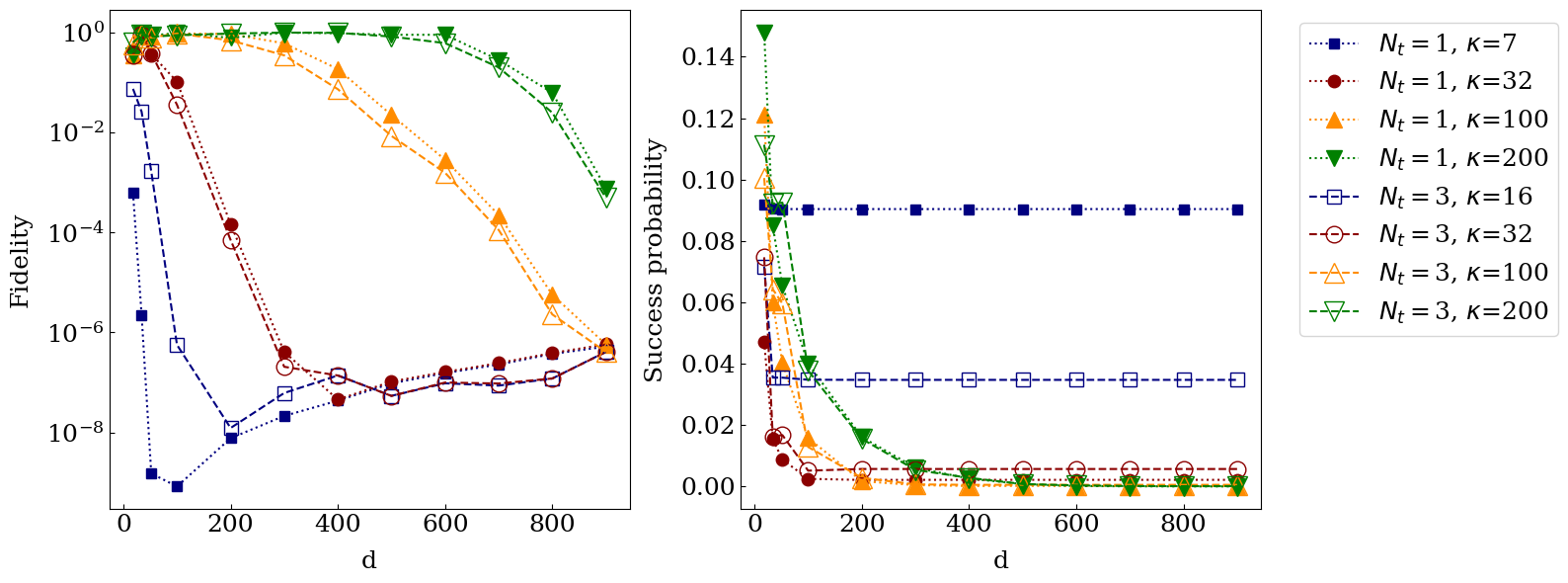} 
\caption{Error fidelity (left panel) and success probability (right panel) of the QSVT solver as a function of the polynomial degree $d$, evaluated under different prescribed condition numbers $\kappa$. Filled symbols correspond to an evolution of $N_t = 1$ ($\kappa_{\rm ex} = 7$), and empty symbols correspond to $N_t = 3$ ($\kappa_{\rm ex} = 16$). }
\label{fig:QSVT_condition}
\end{figure*}

In the QSVT linear systems algorithm, the scalar inversion function $1/x$ is approximated by a polynomial of degree $d$ over the domain $\left[\frac{1}{\kappa}, 1\right]$, where $\kappa$ represents a chosen parameter that should ideally track the true condition number $\kappa_{\rm ex}$ of the system matrix $A$. In practice, determining $\kappa_{\rm ex}$ exactly can be computationally expensive, often requiring a lower-bound estimation of the minimum singular value (or the lowest energy eigenvalue of the embedded Hamiltonian). Here, we analyze the performance of the QSVT-FDTD solver when $\kappa_{\rm ex}$ is overestimated by a chosen parameter $\kappa$.\\
For this evaluation, we implement the QSVT method on the benchmark system described in Sec.~\ref{sec:1D_case} while varying the prescribed parameter $\kappa$. The left and right panels of Fig.~\ref{fig:QSVT_condition} display the resulting error fidelity and success probability, respectively, as a function of the polynomial degree $d$. The various geometric symbols distinguish different values of $\kappa$, where filled symbols correspond to the $N_t = 1$ case (with matrix condition number $\kappa_{\rm ex} = 7$) and empty symbols correspond to the $N_t = 3$ case (with $\kappa_{\rm ex}= 16$).\\
We observe that when $\kappa$ closely matches the true condition number $\kappa_{\rm ex}$, the algorithm achieves rapid convergence with a negligible fidelity error. Conversely, a mismatch between $\kappa_{\rm ex}$ and $\kappa$ severely degrades performance, requiring a significantly larger polynomial degree $d$ to restore the solution accuracy to the asymptotic regime. \\
Crucially, we highlight that this sensitivity is highly dependent on the choice of the phase-factor optimization protocol. In this work, the QSVT angles are synthesized using the \texttt{qsppack} package~\cite{qsppack_github}; alternative optimization methodologies or optimization landscapes may exhibit different resilience profiles to condition number bounds.

 \end{document}